\documentclass[journal]{IEEEtran}
\usepackage{microtype} 
\usepackage{multirow}
\usepackage{graphicx}
\usepackage{booktabs}
\usepackage{amsmath}
\usepackage{mathtools}
\usepackage{hyperref}
\usepackage{balance}
\usepackage{url}            % simple URL typesetting
\usepackage{amsfonts}       % blackboard math symbols
\usepackage{nicefrac}       % compact symbols for 1/2, etc.
\usepackage{xcolor}         % colors
\usepackage{wrapfig}
\usepackage{enumitem}
\usepackage{float}
\usepackage{physics}
\usepackage{subfigure}
\usepackage{bm}
\usepackage{bbm}
\usepackage{textcomp}
\usepackage{tikz}
\usepackage{amsthm}
\usepackage{thmtools}
\usepackage{thm-restate}
\usepackage{amssymb}
\usepackage{qcircuit}
\usepackage{array}
\usepackage[algo2e, linesnumbered, ruled, vlined]{algorithm2e}
\usepackage{etoolbox}

\newtheorem{proposition}{Proposition}
\newtheorem{theorem}{Theorem}

\newtheorem{corollary}{Corollary}
\declaretheorem[name=Lemma]{lemma}
\newtheorem{assumption}{Assumption}

\newcommand{\bfx}{\ensuremath{\bm {x}}}
\newcommand{\bfy}{\ensuremath{\bm {y}}}

\newcommand{\lcal}{\ensuremath{\mathcal{L}}}
\newcommand{\mcal}{\ensuremath{\mathcal{M}}}
\newcommand{\fcal}{\ensuremath{\mathcal{F}}}

\newcommand{\psib}{\ensuremath{\bm{\psi}}}

\newcommand{\thetab}{\ensuremath{\bm{\theta}}}

\newcommand{\BfPara}[1]{{\noindent\bf#1.}\xspace}

\usepackage[normalem]{ulem}

\newcommand{\tred}[1]{\textcolor{red}{#1}}

\newcommand{\leqwhy}[1]{\stackrel{(\text{#1})}{\leq}}

\begin{document}
\title{Quantum Federated Learning with Entanglement Controlled Circuits and Superposition Coding} %
\author{Won Joon Yun, Jae Pyoung Kim, Hankyul Baek, Soyi Jung,~\IEEEmembership{Member,~IEEE,} Jihong Park,~\IEEEmembership{Senior Member,~IEEE,} Mehdi Bennis,~\IEEEmembership{Fellow,~IEEE,} and Joongheon Kim,~\IEEEmembership{Senior Member,~IEEE}
\thanks{The parts of this research were presented at IEEE Conference on Computer Communications (INFOCOM), London, United Kingdom, May 2022~\cite{infocom2022baek}.}
\thanks{This research is supported by the National Research Foundation of Korea (2021R1A4A1030775) and the Institute of Information \& Communications Technology Planning \& Evaluation (IITP) grant funded by the Korea government (MSIT) (2021-0-00467, Intelligent 6G Wireless Access System). \textit{(Corresponding authors: Soyi Jung, Jihong Park, Joongheon Kim)}}
\thanks{W. J. Yun, J. P. Kim, H. Baek, and J. Kim are with the School of Electrical Engineering, Korea University, Seoul 02841, Republic of Korea (e-mails: \{ywjoon95,paulkim436,67back,joongheon\}@korea.ac.kr).}
\thanks{S. Jung is with the Department of Electrical and Computer Engineering, Ajou University, Suwon 16499, Republic of Korea (e-mail: sjung@ajou.ac.kr).}
\thanks{J. Park is with the School of Information Technology, Deakin University, Geelong, VIC 3220, Australia (e-mail: jihong.park@deakin.edu.au).}
\thanks{M. Bennis is with the Centre for Wireless Communications, University of Oulu, Oulu 90014, Finland (e-mail: mehdi.bennis@oulu.fi).} % .
}
\maketitle

\begin{abstract}
While witnessing the noisy intermediate-scale quantum (NISQ) era and beyond, quantum federated learning (QFL) has recently become an emerging field of study. In QFL, each quantum computer or device locally trains its quantum neural network (QNN) with trainable gates, and communicates only these gate parameters over classical channels, without costly quantum communications. Towards enabling QFL under various channel conditions, in this article we develop a depth-controllable architecture of entangled slimmable quantum neural networks (eSQNNs), and propose an entangled slimmable QFL (eSQFL) that communicates the superposition-coded parameters of eSQNNs. Compared to the existing depth-fixed QNNs, training the depth-controllable eSQNN architecture is more challenging due to high entanglement entropy and inter-depth interference, which are mitigated by introducing entanglement controlled universal (CU) gates and an inplace fidelity distillation (IPFD) regularizer penalizing inter-depth quantum state differences, respectively. Furthermore, we optimize the superposition coding power allocation by deriving and minimizing the convergence bound of eSQFL. In an image classification task, extensive simulations corroborate the effectiveness of eSQFL in terms of prediction accuracy, fidelity, and entropy compared to Vanilla QFL as well as under different channel conditions and various data distributions.

% As the beyond of noise intetlyrmediate-scale quantum (NISQ) devices are discussed, large-scale quantum systems such as quantum federated learning (QFL) are needed, considering communication and energy heterogeneity.
% In federated learning (FL), the aforementioned requirements can be achieved by slimmable FL (SFL) in classic computers. Therefore, we aim to expand SFL into the realm of quantum computing. However, the process of quantum conversion is not straightforward for the following reasons. The first challenge is that a quantum neural network (QNN) is prohibited from expanding its dimensions. Additionally, QNNs are especially vulnerable to barren plateau, hindering the training process. Finally, the training algorithm for dynamic QNNs has not been discussed. 
% First, to tackle these problems, we propose an entangled slimmable QNN (eSQNN) that can control its depth flexibly and alleviate barren plateaus.
% Secondly, inplace fidelity distillation (IPFD) regularizer is implemented to ensure that local training is successful at any depth of eSQNN. 
% Finally, we propose an entanglement slimmable QFL (eSQFL) that leverages superposition coding and successive decoding in FL.
% Furthermore, we provide the convergence analysis, insights, and methods to obtain the minimal bound. 
% The results of extensive experiments corroborate the advantages of eSQNN, IPFD, and eSQFL.
\end{abstract} 

\begin{IEEEkeywords}Quantum Machine Learning, Quantum Entanglement, Quantum Federated Learning, Superposition Coding
\end{IEEEkeywords}

\section{Introduction}\label{sec:1}

\subsection{Background and Motivation}
Recent advances in quantum computing hardware and algorithms have recently lead to the emergence of quantum machine learning (ML)~\cite{arute2019quantum,AAAI23Wonjoon,
icdcs2022yun}. As opposed to classical computation at a linear scale in bits, quantum computing can perform calculations at an exponential scale in qubits \cite{Shor97}. The main enablers are the stochastic nature and the entanglement phenomenon of qubits, allowing one to make each qubit represent superimposed multiple states and to simultaneously control multiple qubits, respectively. Consequently, even in the current era of noisy intermediate scale quantum (NISQ) \cite{Preskill2018NISQ}, \textit{i.e.}, with 50 to a few hundred qubits, quantum ML has achieved linear or sublinear complexity in various applications, as compared with the polynomial complexity of classical~ML~\cite{chen20}. 

\begin{figure}\centering
    \subfigure[Vanilla QFL.]{\includegraphics[width=.9\columnwidth]{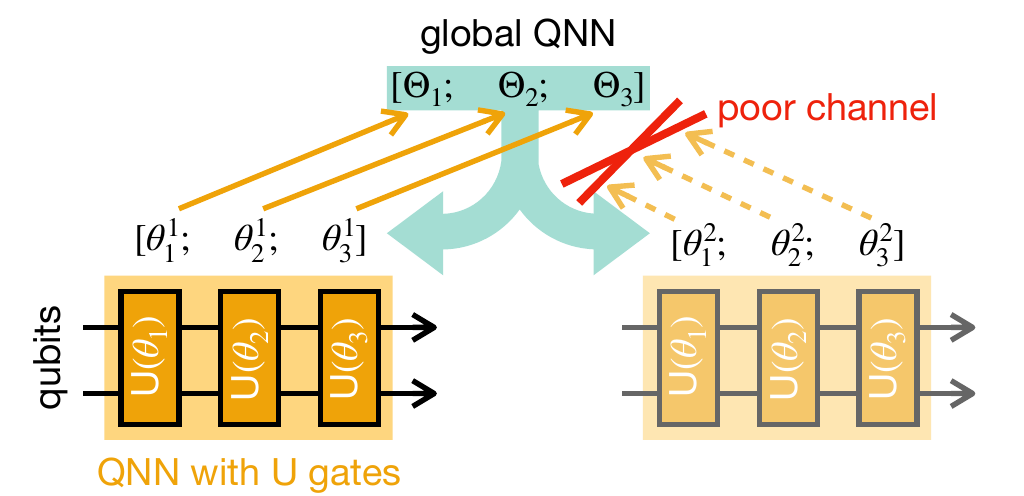}}\\\vspace{-2mm}
    \centering
    \subfigure[eSQFL.]{\includegraphics[width=.9\columnwidth]{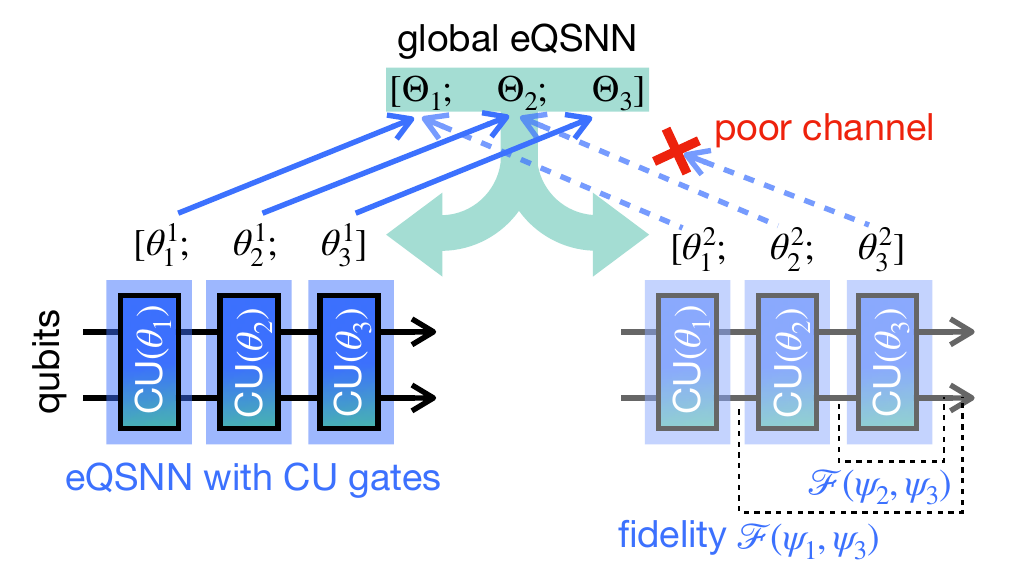}}
     \caption{A schematic illustration of (a) Vanilla quantum federated learning (QFL) and (b) the proposed \emph{entangled slimmable quantum FL (eSQFL)} with $2$ devices each of which has the \emph{entangled slimmable quantum neural network (eSQNN)} having $3$ depth layers.}
     %\vspace{-5.5mm}
    \label{fig:abstract}
\end{figure}

Quantum ML has recently established its standard framework. As analogous to the neural network (NN) of classical ML, the parameterized quantum circuit (PQC), also known as the quantum NN (QNN), has become a de facto standard quantum ML architecture \cite{chen20,KWAK2022}. In a PQC, qubits flow through the gates associated with trainable classical parameters, during which the states of the qubits can be adjusted. For various applications ranging from image classification~\cite{havlivcek2019supervised} to reinforcement learning~\cite{lockwood2020reinforcement}, with a much smaller number of trainable parameters, PQC training has achieved the prediction accuracy on par with the neural network (NN) of classical ML. 

Focusing on the parameter efficiency of PQCs, by integrating federated learning (FL)~\cite{8770530,pieee202105park,9141214,9067847} into standalone quantum ML, quantum FL (QFL) has recently attracted attention~\cite{chen2021federated,tonquant1}. 
Without communicating qubits via costly quantum communications, QFL enables distributed quantum ML at scale by communicating the PQC's trainable parameters via classical communications, even over wireless channels~\cite{quantumfed}. This is not in the distant future, but is an upcoming application, especially considering the ever-increasing pace of innovation in quantum computers, \textit{e.g.}, IBM's development roadmap planning to implement a 1K-qubit beyond-NISQ computer in 2023~\cite{cho2020ibm} and a 100K-qubit computer in 2026~\cite{roadmap2022}.

% \begin{figure}
%     \centering
%     \begin{tabular}{@{}c@{}c@{}}
%     \includegraphics[width=.95\columnwidth]{Figures/Fig_VanillaQFL.pdf} & \includegraphics[width=.95\columnwidth]{Figures/Fig_eSQFL.pdf}\\
%     \small (a) Vanilla QFL. &
%     \small (b) eSQFL. \\
%     \end{tabular}
%      \caption{A schematic illustration of (a) vanilla quantum federated learning (QFL) and (b) the proposed \emph{entangled slimmable quantum FL (eSQFL)} with $2$ devices each of which has the \emph{entangled slimmable quantum neural network (eSQNN)} having $3$ depth layers.}
%     \label{fig:abstract}
% \end{figure}

\subsection{Algorithm Design Concept}
Motivated by this trend in QFL, the overarching goal of this article is to develop a communication-efficient QFL framework that can cope with heterogeneous and time-varying channel conditions and computing resources. To this end, we first revisit slimmable FL (SFL) in classical ML \cite{infocom2022baek}, wherein each device has a width-controllable local model, known as a \emph{slimmable NN (SNN)} \cite{ICCV2019_USlimmable,ijcnn2019kim}, and communicates its superposition-coded local model with different width levels, enabling multi-level local information exchanges depending on channel conditions. Inspired from this, as visualized in Fig.~\ref{fig:abstract}, we propose an \emph{entangled slimmable quantum FL (eSQFL)} framework with \emph{entangled slimmable QNNs (eSQNNs)}, which is a non-trivial extension from SFL with SNNs to their quantum versions as summarized later.

Unlike multi-width SNNs, the eSQNN is a multi-depth PQC wherein more depth levels incur higher \emph{von Neumann entanglement entropy} on average. Unfortunately, the PQC trainability is often challenged by the problem of vanishing all gradients, known as the barren plateaus~\cite{mcclean2018barren}, which is exacerbated under higher entanglement entropy~\cite{sack2022avoiding}. Meanwhile, too low entanglement may negate the benefit of quantum ML. To resolve this issue for an unknown target degree of entanglement, our proposed eSQNN entangles different qubits using the \emph{controlled universal (CU) quantum gates}~\cite{sleator1995realizable} such that the degree of entanglement is trainable.

Next, simultaneous local training of the multiple eSQNN depths may induce inter-depth interference, hindering convergence. In classical ML, SFL avoids its similar inter-width interference issue by adding the inplace knowledge distillation (IPKD) regularizer that penalizes the output difference from any smaller width to the largest width level~\cite{ICCV2019_USlimmable}. Since the IPKD uses the Kullback-Leibler (KL) divergence, it becomes less accurate (or even diverging) for the larger differences. Alternatively, leveraging the \emph{Uhlmann's fidelity} function in quantum information theory~\cite{wilde2013quantum}, we propose a novel \emph{inplace fidelity distillation (IPFD)} regularizer that is bounded within 0 and 1 while accurately measuring the quantum state difference even between the smallest and the largest levels. 

Finally, for communication efficiency, the eSQNN parameters in multiple depths are superposition-coded and transmitted with a different transmit power allocation to each depth. Like SFL, the transmit power is optimized by deriving and minimizing the convergence bound of the eSQFL. Nevertheless, the convergence analysis is completely different since the gradient in PQC training is measured in a quantum computing way using the \emph{parameter shift rule}~\cite{mitarai18}.

Not only by analysis but also by extensive simulations, we corroborate that the proposed eSQFL with eSQNNs achieves convergence while each different width level can be trained to be a separate model with reasonable accuracy and fidelity, under various channel conditions as well as independent and identically distributed (IID) or non-IID data distributions. Note that unlike eSQFL, Vanilla quantum FL having fixed local PQC architectures cannot cope with different channel conditions \cite{chen2021federated, chehimi2022quantum}. A recent work \cite{icml-dynn2022yun} also considers a slimmable architecture in the context of QFL. However, it does not theoretically guarantee convergence, and its specific architecture (\textit{i.e.}, angle/pole parameters) only allows two-level superposition coding, as opposed to generalized multi-level architectures using CU gates in eSQFL.

\subsection{Contributions}
The major contributions of the work in this paper can be summarized as follows.
% \tblue{The proposed eSQFL framework is summarized by Fig.~\ref{fig:abstract}, and the contributions of this paper are four-folded:}
\begin{itemize} %[leftmargin=10pt]
    \item A multi-depth QNN architecture with CU gates, \textit{i.e.}, eSQNN, is proposed to enable superposition-coded transmissions while avoiding barren plateaus. We measure von Neumann entropy between inter-quantum states of different depths. Indeed, CU gates increase the trainability when designing multi-depth QNN.
    
    \item A local eSQNN training algorithm with a fidelity-inspired regularizer, \textit{i.e.}, IPFD, is proposed in order to mitigate inter-depth interference. In eSQNN training, the proposed IPFD in this paper shows the crucial role.
    
    \item With eSQNNs and IPFD, a novel quantum FL framework, \textit{i.e.}, eSQFL, is proposed, and its convergence bound is theoretically derived.
    
    \item Based on the derived convergence bound, transmit power allocation in superposition coding is optimized. In addition, we corroborate that the derived convergence bound helps eSQFL achieve high accuracy.
\end{itemize}

\subsection{Organization} 
The rest of this paper is organized as follows.
Sec.~\ref{sec:2} presents the related work to the proposed quantum federated learning.
Sec.~\ref{sec:3} introduces the eSQNN architecture and its local training with IPFD regularizer. 
Sec.~\ref{sec:4} describes superposition coding, successive decoding, and the proposed eSQFL framework.
Sec.~\ref{sec:5} provides the convergence analysis on eSQFL and its insight.
Sec.~\ref{sec:6} presents the numerical experimental results to corroborate eSQFL empirically.
Lastly, Sec.~\ref{sec:7} concludes this paper.
Notice that the notations in this paper are in Tab.~\ref{tab:notations}.

%More related contributions are in Sec.~\ref{sec:2-1}.

% contributions of this paper are listed below.five-folded: First, a novel dynamic QNN (named eSQNN) that works in multiple depth configurations to avoid barren plateaus is proposed.
% Second, a novel local training algorithm of eSQNN is developed. In addition, IPFD regularization to train all the depth configurations of eSQNN is proposed as well. 
% Third, the first-ever quantum version of SlimFL (named eSQFL) by combining eSQNN and its training method is proposed. Fourth, the convergence of eSQFL, which leverages the characteristic of PQC, is proven. Additionally, the optimal power allocations regarding eSQFL are provided. Finally, our analysis, eSQNN, and IPFD are corroborated via extensive experiments. From these results, our proposed QFL is shown to achieve higher accuracy in various channel quality and data distributions.

% numerical results, ...vs. FL with classical neural network with the same number of parameters, baseline quantum FL without slimmable architecture, under IID, non-IID + etc

% QFL
% SlimFL

\section{Related Work}\label{sec:2}
\subsection{Quantum Machine Learning Basics}\label{sec:2-1}

\BfPara{Basic Quantum Gates} A qubit is a quantum computing unit where the quantum state is represented with two bases $|0\rangle$, and $|1\rangle$ in Bloch sphere~\cite{bouwmeester2000physics}. 
Consider the $q$ qubits system, in which the quantum state defined in Hilbert space $\psib \in \mathbb{C}^{2^q}$ can be expressed as follows,
\begin{equation}
    |\psib\rangle=\Lambda_{1}|0\cdots 0\rangle + \cdots + \Lambda_{2^q}|1\cdots 1\rangle
\end{equation}
where $\sum^{2^q}_{i=1}\Lambda_{i}^2=1$. 
A classical data $\bfx$ is encoded as a quantum state with the rotation gates $R_{\text{x}}(\bfx)$, $R_{\text{y}}(\bfx)$, and $R_{\text{z}}(\bfx)$, where the rotation of $(\bfx)$ occurs in the direction of $x$-, $y$-, and $z$-axes in Bloch sphere, respectively. 
Moreover, qubits are entangled with \textit{controlled-NOT} gates (CNOT)~\cite{williams1998explorations}. CNOT gates act on two qubits to entangle them by using the first qubit as the control qubit and performing \textit{XOR} operation on the second qubit.
These basic quantum gates configure the QNNs. 

\BfPara{Quantum Neural Network} 
%QNNs mimic the \textit{de facto} NNs. 
The structure of a QNN is tripartite: the state encoder, PQC, and the measurement layer~\cite{killoran2019continuous,SIG-118}.
In the forward propagation, classical input data $\bfx$ needs to be first encoded with the state encoder via basic rotation gates, which is a unitary operation and denoted as $U(\bfx)$. Then, the encoded quantum state is processed through the PQC $U(\mathbf{\thetab})$, a multi-layered set of CNOT gates and rotation gates associated with trainable parameters $\thetab$. The quantum state $\psib$ can be expressed as,
\begin{equation}
    |\psib_{\thetab}\rangle = U(\thetab)\cdot|\psib_0\rangle = U(\thetab) \cdot U(\bfx)|0\rangle.
\end{equation}
%$|\psib_{\thetab}\rangle = U(\thetab)\cdot|\psib_0\rangle = U(\thetab) \cdot U(\bfx)|0\rangle$

The output of the PQC is the entangled quantum state that can be measured after applying a projection matrix $M \in \mcal \equiv \{M_1,\cdots, M_c, \cdots, M_C\}$ onto the reference $z$-axis. 
The measured output $\langle V \rangle_{\thetab} \in [-1,1]^{C}$ is called an \textit{observable}, where $C$ denotes the output dimension. 
The operation of QNN corresponding to $c$-th observable is as follows,
\begin{equation}\label{eq:psr-2}
\langle V_c \rangle_{\thetab} =  \langle0|U^{\dagger}(\bfx)U^{\dagger}({\thetab})M_c U({\thetab})U(\bfx)|0\rangle =\langle\psib|M_c|\psib\rangle
\end{equation} 
where $(\cdot)^\dagger$ denotes the complex conjugate operator. Using the observable, a given loss function is calculated. Unlike classical NNs having visible activations in their hidden layers, the quantum states within QNNs are not measurable; otherwise, the quantum states collapse \cite{bouwmeester2000physics}. This does not allows quantum ML to compute the loss gradients via the chain rule, i.e., backpropagations. Alternatively, quantum ML evaluates the gradients using the zero-th order method called the parameter shift rule \cite{mitarai18} (see Appendix~\ref{sec:parameter-shift}).

\subsection{Classical Federated Learning}\label{sec:2-2}
Federated learning (FL) is a machine learning (ML) architecture made up of a server, local devices, and a global model~\cite{pieee202105park}. The server transmits the global model to all the local devices, and each device produces local parameters by training the received global model. Then, these parameters are sent back to the central server, where all the data is aggregated to update the global model. Finally, the updated global model is transmitted to the local devices again for another iteration. Due to this mechanism, FL allows a large number of devices to learn a global model simultaneously without transmitting any data, ensuring data privacy as well. Considering the recent increase in the number and computational power of edge devices, FL is an extremely useful tool for reducing computational overhead and protecting data security which is both emerging challenges in the field of ML~\cite{TranBZMH19}.
Within this architecture, various techniques with differing methods of aggregating data and training the global model exist, \textit{e.g.}, FedAvg~\cite{mcmahan2017communication}, FedBN~\cite{li2021fedbn}.

The convergence analysis of FL algorithms is especially challenging because of the data heterogeneity in FL, which forces researchers to rely on copious numbers of assumptions. Consequently, gaps in the understanding of FL analysis occur.
Over the years, many major works have attempted to better understand FL by removing assumptions and exploring new techniques~\cite{mangasarian1995parallel, cotter2011better, mcmahan2017communication,ToN3,tonfl1}. Even now, research on FL convergence in various aspects is still being carried out (\textit{i.e.}, non-convex, convergence bounds). 
For example, \cite{Khaled2020Tighter} has successfully proposed an analysis of local stochastic gradient descent (SGD) using only arbitrarily heterogeneous data while also using weaker assumptions than previous works.
On the other hand, convergence analysis of most QFL algorithms has not been fully developed yet. 
This paper aims to further discuss the convergence analysis of QFL via the analysis of eSQFL with the characteristic of quantum computing. The convergence analysis on a dynamic QFL is elaborated in Sec.~\ref{sec:4}. 

\subsection{Classical Slimmable Federated Learning}\label{sec:2-3}
SFL is a framework that executes FL by using \textit{slimmable neural networks (SNNs)} with \textit{SC} and \textit{SD}~\cite{infocom2022baek}. The architectural properties of SNN reduce memory costs of SFL~\cite{ICCV2019_USlimmable} while with rigorous communication and computational efficiencies. SC is a process of compressing two different data signals into one signal. As the signals are encoded, different power levels are assigned to each data signal which is used to decide the priority of signals during SD.
Additionally, the SNN is composed of the left-hand (LH) and the right-hand (RH) sides. The LH side is occupied by the high priority signal, while the lower priority signal goes to the RH side.
After SC is finished, the encoded message will be uploaded to the server, which then undergoes SD.
Assuming that the state of the communication channel is good, the LH of the SNN will be decoded first, followed by the RH signal. However, if the communication channel is not stable enough, only LH will be decoded, resulting in a small model. Finally, if the communication channel is completely unstable, no signal will be obtained. This flexible characteristic of SNN allows SFL to be extremely adaptable to dynamic communication environments, making it suitable for practical applications.

\subsection{Quantum Federated Learning} 
In this section, the concept of QFL is elaborated in depth. QFL is implementing FL via quantum computation by replacing all the NNs with QNNs. Chen \textit{et al.},~\cite{chen2021federated} is the first to propose a hybrid quantum-classical QFL architecture where the local devices are replaced with quantum devices, unlike FL models. After receiving global model parameters, the quantum computers carry out QML using QNNs. Then, the output of each device is aggregated to update the global model before repeating the process. In Chehimi \textit{et al.},~\cite{chehimi2022quantum}, a purely QFL framework is proposed. Similar to~\cite{chen2021federated}, this model is composed of a server and multiple quantum devices. However, instead of converting classic data into quantum states, the local devices generate quantum data by labeling qubits as excited or not excited according to the degree of rotation on the Bloch sphere. Both \cite{chen2021federated, chehimi2022quantum} use FedAvg to aggregate data and execute training. As seen from the two examples above, a QFL and FL share an identical system structure, but QFL leverages QML instead of ML in order to exploit the advantages of quantum computing. For this work, Vanilla QFL is referring to a purely quantum version of \cite{chen2021federated}. 
In addition, quantum application of SFL is studied to improve the communication opportunities~\cite{icml-dynn2022yun}. 
SQFL utilized trainable measurement parameters to configure two messages which contain both the trainable measurement parameters and PQC parameters, respectively. However, in this work, multiple layer architectures and local training algorithms are proposed, which are not present in \cite{icml-dynn2022yun}.

\section{Architecture and Training of eSQNNs}\label{sec:3}
\begin{figure}[t!]
\centering
\includegraphics[width=\linewidth]{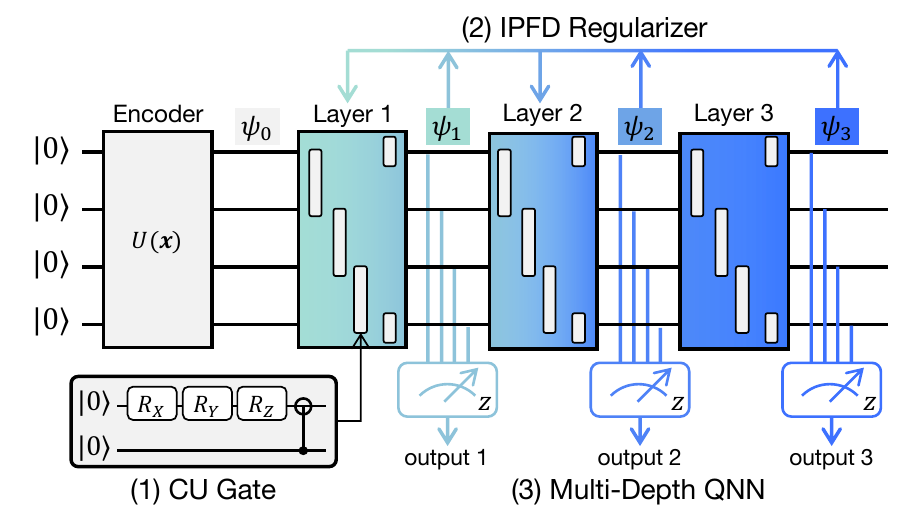}
\caption{Illustration of eSQNN: (1) we use \textit{CU gates} in eSQNN, (2) the \textit{fidelity regularizer} is applied sub-model layers (\textit{e.g.}, \textit{Layer 1 and 2}) by receiving quantum states, (3) our proposed eSQNN is based on the \textit{multi-depth} QNN.}\label{fig:eSQNN}
\end{figure}
In this section, we describe the architecture of eSQNN and its local training algorithm. To elaborate on this, by slightly modifying the depth-fixed architecture of Vanilla QNN \cite{chen20}, we first prepare its depth-controllable counterpart without controlling the level of entanglement, dubbed Vanilla SQNN, followed by introducing the proposed eSQNN controlling both the depth and the level of entanglement.
% because dynamic width and kernel size cannot be used because the number of qubits cannot be reduced or increased. 

\BfPara{Architecture of Vanilla SQNN} 
Suppose that Vanilla SQNN consists of $L$ layers, and produces the desired outputs at any layer $l \in [1, L]$. 
In this paper, the number of sub-models must be larger than 1 (\textit{i.e.}, $L \geq 2$).
When the $l$-th sub-model is used, it means that the $l$-th model will be configured from the encoding layer to the $l$-th layer. 
For an arbitrary $k\in\mathbb{N}[1,K]$ and $l\in\mathbb{N}[1,L]$, the model parameters of $k$-th local device and the $l$-th layer is denoted as $\thetab^k \odot \sum^{l}_{l'=1}\Xi_{l'}$. 
Note that $\Xi_{l'}$ is a binary mask which eliminates all trainable parameters except  parameters of the $l'$-th layer. 
The operation $\odot$ denotes an element-wise product. 
%The $\psib_l$ is denoted as the quantum state.
However, it is difficult to make desirable results at any random layer because the vanilla SQNN is vulnerable to the barren plateau problem \cite{mcclean2018barren,you2021exponential}. 
The barren plateau is a bad local optimum which hinders convergence. 
It is known that more entanglement's degree introduces worse the barren plateau problem~\cite{sack2022avoiding}. 
The operations in Vanilla SQNN are as follows: 1) rotate quantum state $|\psib\rangle$ with rotation gates, 2) entangle qubits, and 3) repeat the first and second steps.
We predict that the operations mentioned above will increase the degree of entanglement.

\BfPara{Architecture of eSQNN}
eSQNN is proposed to cope with the problem of Vanilla SQNN architecture. Fig.~\ref{fig:eSQNN} shows the illustration of eSQNN. 
%It can be seen that each eSQNN $\thetab^k$ is subdivided into $L$ sub-models. 
eSQNN is mainly composed of CU gates. 
The operations of the CU gate in two qubits are written as 
$\begin{bmatrix}
    I & 0\\
    0 & U
    \end{bmatrix}$, 
where $U$ is expressed as 
$U=\begin{bmatrix}
    u_{00} & u_{01}\\
    u_{10} & u_{11}
\end{bmatrix}$. 
Note that $U$ is an unitary matrix, \textit{i.e.}, $U^\dagger U = I$. 
We focus on the architectural advantage of CU gates because CU gates can adjust the direction of entanglement, disentanglement, or rotation while training. We describe the advantages of eSQNN and the barren plateau phenomenon next.

To this end, at first we consider the \textit{von Neumann entanglement entropy}, a metric for measuring the degree of quantum entanglement of bipartite subsystems in an entire system~\cite{greenberger2009compendium}. For instance, consider two subsystems, \textit{e.g.}, $l$ and $l'$-th model configuration, where $l > l'$. 
According to a two-copy test from \cite{Suba__2019}, we can compare the different quantum states $|\psib_l\rangle$ and $|\psib_{l'}\rangle$ by using additional qubits. 
Then, we can measure the entanglement entropy by following the statement below.
Suppose a quantum state that exists in $l$ and $l'$-th depth is represented as 
$\psib_{l',l} \in \mathbb{C}^{2^{2q}}$. 
Its pure state is obtained by $\rho_{l',l} \triangleq |\psib_{l',l}\rangle\langle\psib_{l',l}|$. Finally, the entanglement entropy is calculated as follows, 
\begin{equation}
    S_{l}(\rho_{l',l}) = -\Tr_{l}(\rho_{l',l} \log\rho_{l',l})
\end{equation}
%$S_{l}(\rho_{l',l}) = -\Tr_{l}(\rho_{l',l} \log\rho_{l',l})$,
where $\Tr_{l}(\cdot)$ stands for partial trace over the  $l$-th layer. 
As discussed in many studies, avoiding the barren plateaus requires the reduction of the entanglement entropy~\cite{sack2022avoiding}. 

On the basis of these studies, we assume that there exists an entropy threshold for every $l$-th model, \textit{i.e.,} $S_{l,th}$ for all $\forall l \in \mathbb{N}[l', L]$ and $\forall l' \in \mathbb{N}[0, l-1]$. It starts at $l'=0$, because we measure the entanglement entropy from the encoding state, \textit{i.e.}, $\psib_0$.
If $\sum^{l-1}_{l'=0}S(\rho_{l',l}) \geq S_{l,th}$, the barren plateau becomes severe and training of $l$-th model fails. 
For this, we observe the entanglement entropy between the encoding state and the layer of eSQNN.
In order to ensure all model configurations are trained, we define a metric as, 
\begin{equation}
\mathbbm{1}_{\text{train}} = \prod^L_{l=1}\mathbbm{1}\left(\sum^{l-1}_{l'=0}S(\rho_{l',l}) < S_{l,th}\right) 
\end{equation}
where $\mathbbm{1}(\cdot)$ stands for an indicator function.
In order to verify whether the metric works correctly, we provide the following two cases. 
If all model configurations are satisfied $\sum^{l-1}_{l'=0}S(\rho_{l',l}) < S_{l,th})$, then $\mathbbm{1}_{\text{train}} = 1$,
%$\sum^{l-1}_{l'=0}S(\rho_{l',l}) < S_{l,th})$, $\mathbbm{1}_{\text{train}} = 1$
% \begin{equation}
%     \sum^{l-1}_{l'=0}S(\rho_{l',l}) < S_{l,th}),~\mathbbm{1}_{\text{train}} = 1
% \end{equation},
which means it avoids the barren plateau. On the other hand, suppose that $\exists l$ that satisfies $\sum^{l-1}_{l'=0}S(\rho_{l',l}) \geq S_{l,th}$, we have $\mathbbm{1}_{\text{train}} = 0$ which it means it suffers from barren plateau.
We conjecture that eSQNN is robust to the barren plateau than Vanilla SQNN because the event $\mathbbm{1}_{\text{train}} = 1$ frequently occurs in eSQNN. More details are in Sec.~\ref{sec:6-2}.

\BfPara{eSQNN Local Training} 
This section presents the eSQNN local training algorithm. In general, classic SNNs use the IPKD regularizer $\lcal_{KD}$ to transfer knowledge from a large model to a small model~\cite{ICCV2019_USlimmable}, which can be expressed as, 
%$\lcal_{KD} = D_{KL}(p(\bfy^{k,L}_{t,e})\|p(\bfy^{k,l}_{t,e}))$
\begin{equation}
    \lcal_{KD} = D_{KL}(p(\bfy^{k,L}_{t,e})\|p(\bfy^{k,l}_{t,e}))
\end{equation}
where $D_{KL}$ is the KL divergence. IPKD is ill-suited when the difference between the outputs of two models becomes large, where the KL divergence may even diverge. Alternatively, we propose the IPFD regularizer $\lcal_{FD}$, inspired by the Uhlmann's fidelity function \cite{jozsa1994fidelity} in quantum information theory, measuring the similarity between two quantum states~\cite{jozsa1994fidelity}.
Precisely, the fidelity of the quantum states in the $L$-th and $l$-th model configurations is defined as follows,
%$\fcal (\psib_L, \psib_l) = |\langle \psib_L| \psib_l\rangle|^2$. 
\begin{equation}
    \fcal (\psib_L, \psib_l) = |\langle \psib_L| \psib_l\rangle|^2.\label{eq:fid}
\end{equation}

In \eqref{eq:fid}, if $\fcal (\psib_L, \psib_l) \approx 1$, $\psib_l$ is similar to $\psib_L$, which means the logits of $l$-th model are almost same as the logits of $L$-th model. On the other hand, the opposite condition $\fcal (\psib_L, \psib_l) \approx 0$ means the $l$-th model does not follow the $L$-th model. 
% Spurred by the definition of fidelity, we design IPFD regularizer, where the $L$-th model teaches the $l$-th model via fidelity. 

Consequently, in a classification task, the local training of an eSQNN with the IPFD regularizer is described as follows. The parameters $(\bfx, \bfy)$ are denoted as data and label, respectively. The predicted label $\bfy =\{y_c\}^C_{c=1}$ is an one-hot encoded vector wherein the element $y_c$ becomes unity for a true label and otherwise $0$, \textit{i.e.}, $y_{c'}=0, \forall c'\neq c$. 
Hereafter, we describe the local training for the parameters of local device $k$ in the $t$-th communication round and $e$-th local training iteration.
The logits of class and its prediction of $l$-th model are denoted as,
\begin{align}
    y^{k,l,c}_{t,e}&=\text{exp}(a \langle V_c \rangle_{\thetab^k_{t,e} \odot \sum^{l}_{l'=1}\Xi_{l'} }), \\
    p(y^{k,l,c}_{t,e}|\bfx) &= \frac{y^{k,l,c}_{t,e}}{\sum^C_{c=1}y^{k,l,c}_{t,e}},
\end{align}
%$y^{k,l,c}_{t,e}=\text{exp}(a \langle V_c \rangle_{\thetab^k_{t,e} \odot \sum^{l}_{l'=1}\Xi_{l'} })$ 
%and $p(y^{k,l,c}_{t,e}|\bfx) = \frac{y^{k,l,c}_{t,e}}{\sum^C_{c=1}y^{k,l,c}_{t,e}}$ respectively. 
where $a$ represents the observable hyperparameter. Additionally, the cross-entropy loss and the fidelity regularization are as,
\begin{align}
    \lcal_{CE} &= - \sum^C_{c=1} [y_c \text{log}(p(y^{k,l,c}_{t,e})|\bfx)],\\ \lcal_{FD} &= 1-\fcal(\psib^{k,L}_{t,e,\bfx},\psib^{k,l}_{t,e,\bfx}).
\end{align}
%$\lcal_{CE} = - \sum^C_{c=1} [y_c \text{log}(p(y^{k,l,c}_{t,e})|\bfx)]$ and $\lcal_{FD} = 1-\fcal(\psib^{k,L}_{t,e,\bfx},\psib^{k,l}_{t,e,\bfx})$ respectively, are used to formulate the loss function as shown below.

The loss function is given as,
\begin{align}
\lcal^{k,l}_{t,e} =\frac{1}{D} \sum_{(\bfx, \bfy) \in \zeta^k}
\left[\lambda\lcal_{CE} +(1-\lambda) \lcal_{FD}\right] \label{eq:loss}
\end{align}
where $D$ and $\lambda$ stand for the batch size and the balanced parameter of fidelity regularization, respectively. 
The gradient of \eqref{eq:loss} can be calculated with parameter shift rule~\cite{mitarai18}.
{Algorithm~\ref{alg:a2p}} summarizes the local training process before one communication round. 
After training with {Algorithm~\ref{alg:a2p}}, the gradient to be transmitted to the server can be as,
\begin{equation}
g^k_t = \sum^{E}_{e=1} \sum^{L}_{l=1} \nabla_{{\thetab^k_{t,e}}} \lcal^{k,l}_{t,e} 
\end{equation}
where $\eta_t$ denotes the learning rate at communication round $t$. 

 \begin{algorithm2e}[t]
% \small
    \SetCustomAlgoRuledWidth{0.44\textwidth}  
\caption{Local-eSQNN Training}\label{alg:a2p} 
\textbf{Initialization.} local-QNN parameters, $\thetab$\;
 \For{ $e = \{1,2,\dots, E\}$}
 {
     \For{ $(\bfx, y) \in \mathcal{D}$}
        {    Get logits with $L$-th model\;
             Calculate loss with labels and accumulate loss\;
             \For{$l = \{1,2,\dots,L-1\}$}
             {  
                Get logits with $l$-th model\;
                Calculate loss gradient with parameter-shift rule\;
                
             }
             $\thetab^k_{t,e+1}\leftarrow \thetab^k_{t,e}-\eta_t\nabla_{\theta^k_{t,e}}\lcal^{k,l}_{t,e}$\;
         }
 }
\end{algorithm2e}

\section{Entangled Slimmable Quantum Federated Learning}\label{sec:4}

\subsection{Superposition Coding \& Successive Decoding}
The successful reception of a wireless signal is mainly affected by the signal-to-interference-plus-noise ratio (SINR) \cite{TseBook:FundamaentalsWC:2005}.
At a receiver, SINR can be expressed as,
\begin{equation}
    \gamma={\chi d^{-\beta} P}/{(\sigma^2 + P^I)}
\end{equation}
%$\gamma={\chi d^{-\beta} P}/{(\sigma^2 + P^I)}$,
where $P$, $P^I$, $d$, and $\sigma^2$ denote the transmission interference, reception interference, a transmitter-receiver distance, and noise powers, respectively. In addition, $\beta\geq 2$ is a path loss exponent and $\chi$ is small-scale fading parameter {(\textit{i.e.}, Rayleigh fading)}. Following the Shannon's capacity formula with a Gaussian codebook, the received throughput $R$ with the bandwidth $W$ is $R = W \log_2(1 + \gamma)$ (bits/sec).
When the transmitter encodes raw data with a code rate~$u$, its receiver successfully decodes the encoded data if $R > u$. Finally, the decoding success probability can be given as follows,
%$\Pr(R \geq u)= \Pr(\frac{\chi d^{-\beta} P }{\sigma^2 + P^I} \geq u')$, where $u'=2^{\frac{u}{W}}-1$. 
\begin{equation}
    \Pr(R \geq u)= \Pr(\frac{\chi d^{-\beta} P }{\sigma^2 + P^I} \geq u')
\end{equation}
where $u'=2^{\frac{u}{W}}-1$.
Consider transmitting $L$ messages from a transmitter to a receiver simultaneously. Before transmission, these messages are SC-encoded \cite{Cover:TIT72}, and the whole transmission power budget $P$ is allotted to the $l$-th message, with $P_l = \nu_l P$ transmission power for $l \in [1,L]$. Note that $\nu_{l}$ is an allocation variable such that $\nu_{l}>u'\sum^L_{l'=l+1}\nu_{l'}$, $\sum^L_{l=1}\nu_l=1$, and $\forall \nu_l \geq 0$.

The SC-encoded message is meant to be sequentially decoded at the receiver by first decoding the strongest signal, then canceling out the decoded signal, and finally decoding the next strongest signal, \textit{i.e.}, SD, also known as successive interference cancellation~\cite{Jinho:TCOM17,8784249}. The small-scale fading parameter $\chi$ under Rayleigh fading follows an exponential distribution, \textit{i.e.}, $\chi\sim\exp(1)$. Assuming $l'>l$, the receiver may gradually decode the $l$-th message while experiencing the remaining messages as interference $P_l^I$, \textit{i.e.}, 
% $P_l^I = \chi d^{-\beta} P \sum^{L}_{l' = l+1}\nu_{l'}$,
\begin{equation}
    P_l^I = \chi d^{-\beta} P \sum^{L}_{l' = l+1}\nu_{l'},
\end{equation}
for $l\leq L-1$. However, $P^I_L=0$ as there is no interference for the last message. Assume that $R_l$ represents the throughput of the $l$-th message. Then, the distribution of $R_l$ is given as, 
% $\Pr(R_l \geq u ) = \text{Pr}(\chi \geq \frac{1/\bar{\gamma}}{ \nu_l/u' - \sum^L_{l'=1+1}\nu_l'}\Big)$,
\begin{equation}
    \Pr(R_l \geq u ) = \text{Pr}\Big(\chi \geq \frac{1/\bar{\gamma}}{\nu_l/u' - \sum^L_{l'=1+1}\nu_l'}\Big)
\end{equation}
where $\bar{\gamma}=\frac{P d^{-\beta}}{\sigma^2}$ denotes the averaged signal-to-noise ratio (SNR).
By using this result, the $l$-th message's decoding success probability $p_l$ can be expressed as follows,
\begin{align}
    p_l&= \Pr(R_1 \geq u, \cdots, R_l \geq u), \\
    = &\text{Pr}\Big(\!\! \chi \geq  \max(
    \frac{1/\bar{\gamma}}{ \frac{\nu_l}{u'} \!-\! \sum^L_{l'=2}\nu_{l'}} ,\! \cdots\!,\!  \frac{1/\bar{\gamma}}{ \frac{\nu_l}{u'} \!-\! \sum_{l'=l+1}\nu_{l'}})\!\!\Big).
\end{align}
% $   p_l= \Pr(R_1 \geq u, \cdots, R_l \geq u) = \text{Pr}\Big( \chi \geq  \max(
%     \frac{1/\bar{\gamma}}{ \nu_1/u' - \sum\nolimits^L_{l'=2}\nu_{l'}} , \cdots,  \frac{1/\bar{\gamma}}{ \nu_l/u' - \sum\nolimits^L_{l'=l+1}\nu_{l'}})\Big)$.

\begin{algorithm2e}[t]
% \small
\caption{eSQFL}\label{alg:sqfl}
\textbf{Notation.} $\thetab^k_t$: $k$-th device's parameters, $\Theta_t$: parameters of global eSQNN, $X_l$: set of $l$-th subdivided gradient\;
\textbf{Initialization.} 
$X_l \leftarrow \emptyset, \forall l \in [1,L]$ \; 
 \For{ $k = \{1, \dots, K\}$}
 {
     Sample $\chi^k \sim \exp(1) $\;
     \For{$l=\{1,2,\dots,L\}$}
     {
         \If{$\chi_k \geq u_{l}$}
         { 
            $X_l \leftarrow X_l \cup k$\;
         }
     }
 }
 \If{$\prod^L_{l=1} \mathbbm{1}(X_l=\emptyset) \neq 0$}
 {
     $\Theta_{t+1} \leftarrow \Theta_{t} -\eta_t  \sum\nolimits^L_{l=1}\frac{1}{|X_l|}\sum\nolimits_{k\in X_l}g^k_t \odot \Xi_l$ \;
 }
 \Else
 {
  Skip aggregation\;
 }
 \For{$k = \{1, \cdots, K\}$}
{
    $\thetab^k_{t+1, 1} \leftarrow \Theta_{t+1}$\;
}
\end{algorithm2e}

\subsection{eSQFL Operations} 
This section describes the operations of eSQFL. 
{Algorithm~\ref{alg:sqfl}} shows the eSQFL algorithm.
First of all, local devices are trained with {Algorithm~\ref{alg:a2p}}. 
The power allocation is conducted to configure SC-encoded model parameters, \textit{i.e.}, $\bm \nu = \{\nu_l\}^L_{l=1}$ for the gradient $\{g^k_t \odot \Xi_l\}^L_{l=1}$ of the subdivided model configuration.
After that, the local devices transmit their SC-encoded model parameters to the server. 
The server decodes the devices' SC-encoded model parameters with SD. 
If the server receives at least one local gradient for every model configuration, the server aggregates; otherwise, no aggregation occurs. In the aggregation of sub-divided model configuration, FedAvg is utilized~\cite{mcmahan2017communication}. The updates of eSQFL will be explained later.

\section{Convergence Analysis}\label{sec:5}
\subsection{Setup}\label{sec:5-1}
In order to analyze the convergence rate of eSQFL, the following assumptions are considered. Firstly, the local-side decoding is always successful ({Algorithm~\ref{alg:sqfl}}, lines 12--13) because the server-side transmission power is higher than the uplink power. Secondly, $K$ is assumed to be big enough such that $|X_l| \approx K p_l $, for all $l$. During the $t$-th communication round, the server builds the global model which can be expressed as follows,
\begin{equation}
    \Theta_{t+1} \leftarrow \Theta_{t}-\eta_t\underbrace{\sum\nolimits^L_{l=1}\frac{1}{K p_l}\sum\nolimits_{k \in X_l} g^{k}_{t} \odot \Xi_l}_{:=f_t}.
    \label{eq:global}
\end{equation}

The objective function of the global model and the local objective functions are denoted as $F$ and $\{F^k\}$ respectively. 
The bar notation $\bar{\cdot}$ is used for the averaged value over $\{\zeta_t^k\}$, and the superscript $^*$ is used to indicate the optimum.
For mathematical amenability, we consider the following assumptions on $F$ and $\{F^k\}$, as used in \cite{li2019convergence}.
\begin{assumption}[\textbf{$\bm \beta$-Smoothness}] \label{assum:1}
If $F$ and $\{F^k\}$ are $\beta$-smooth,  
	% $F^k(\thetab_{v})  \leq F^k(\thetab_{w}) + (\thetab_{v} - \thetab_{w})^T \nabla F^k(\thetab_{w}) + \frac{\beta}{2} \| \thetab_{v} - \thetab_{w}\|^2$ 
 \begin{multline}
     F^k(\thetab_{v})  \leq \\ F^k(\thetab_{w}) + (\thetab_{v} - \thetab_{w})^T \nabla F^k(\thetab_{w}) + \frac{\beta}{2} \| \thetab_{v} - \thetab_{w}\|^2,
 \end{multline}
	for all $v,w>0$.
\end{assumption}
\begin{assumption}[\textbf{$\bm \mu$-Strong Convexity}] \label{assum:2}
If $F$ and $\{F^k\}$ are $\mu$-strong convex,
	%\textit{i.e.}, $F^k(\thetab_{v})  \geq F^k(\thetab_{w}) + (\thetab_{v} - \thetab_{w})^T \nabla F^k(\thetab_{w}) + \frac{\mu}{2} \| \thetab_{v} - \thetab_{w}\|^2$
    \begin{multline}
        F^k(\thetab_{v})  \geq \\ F^k(\thetab_{w}) + (\thetab_{v} - \thetab_{w})^T \nabla F^k(\thetab_{w}) + \frac{\mu}{2} \| \thetab_{v} - \thetab_{w}\|^2,
    \end{multline}
for all $v, w>0$. 
\end{assumption}
\begin{assumption}[\textbf{Bounded Local Gradient Variance}]\label{assum:3}  
For all device $k \in \mathbb{N}[1,K]$ and its local data $\zeta^k \in \mathbf{Z}$, the difference between the local gradient $F^k(\thetab^k;\zeta^k)$ and $\bar{F}^k(\thetab;\mathbf{Z})$ is bounded, \textit{i.e.,}
\begin{equation}
    \mathbb{E}[\|\nabla_{\thetab}F^k(\thetab^k,\zeta^k_t)- \nabla_{\thetab} \bar{F}^k(\thetab^k;\mathbf{Z})\|^2] \leq \sigma_k^2.
\end{equation}
%$\mathbb{E}[\|\nabla_{\thetab}F^k(\thetab^k,\zeta^k_t)- \nabla_{\thetab} \bar{F}^k(\thetab^k;\mathbf{Z})\|^2] \leq \sigma_k^2$.
\end{assumption}
According to \cite{Khaled2020Tighter}, the metric for the non-IIDness of $\mathbf{Z}$ is given as follows,
\begin{equation}
    \delta  =  \frac{1}{K}\sum^K_{k=1}\sigma_k^2.
\end{equation}
  %\delta  =  \frac{1}{K}\sum^K_{k=1}\nolimits{\sigma_k^2}$.
  
\subsection{Convergence Analysis}\label{sec:5-2}

In classical ML, the convergence of FedAvg has been analyzed by assuming bounded local gradients in~\cite{li2019convergence}. Without such an unrealistic assumption, the convergence bound of SFL has been derived in \cite{infocom2022baek}. In quantum ML, local gradients can be shown to be inherently bounded thanks to the bounded fidelity and the parameter shift rule computing quantum gradients~\cite{mitarai18}. Hence, rather than adopting the methods in \cite{infocom2022baek}, we first derive the local gradient bound, and then derive the convergence bound of eSQFL by following the steps~\cite{li2019convergence}. The detailed proofs are deferred to Appendix, and only the results are presented as elaborated next.

% To analyze the convergence of FedAvg with a classic neural network, the following literature assumes that the local gradient is bounded~\cite{li2019convergence}.
% On the other hand, the local gradient of a QNN is bounded by a constant due to the characteristic of unitary gate operations and parameter-shift rule (refer to Appendix~\ref{sec:parameter-shift})~\cite{mitarai18}.
% Our approach is to derive the bound of the gradient of the global and local models.
% The convergence proof is in Appendix.
% \begin{lemma}[\textbf{Bounding the local gradient}] \label{lemma:2}
% According to the parameter shift rule, the $k$-th device's gradient $\lcal^{k,l}_{t,e}$ over its parameter $\thetab^k_{t,e}$ is bounded as
% $\|\nabla_{\thetab^k_{t,e} \odot \sum^l_{l'=1}\Xi_{l'}} \lcal^{k,l'}_{t,e} \| \leq B_1$, where $B_1 = \frac{a^l (C-1)}{D}
% \sum\limits_{(\bfx,\bfy) \in \zeta^k} \frac{ (\max_{c'} y^{k,l,c'}_{t,e})^2}{\min_{c'} y^{k,l,c'}_{t,e} } 
% \cdot \Big( C^2 + (C-2) \cdot \frac{\sum\limits^{C}_{c=1} y^{k,l+1,c}_{t,e}}{\max_{c'} y^{k,l+1,c'}_{t,e}}\Big)$.
% \end{lemma}
\begin{lemma}[\textbf{Bounded Local Gradient}]\label{lemma:1}
For $ t\geq 1$ and $\eta_t\leq \eta_{t+1}$, it follows that 
%$\mathbb{E}[\|g^k_t\|^2] \leq EL (2+(a-2)\lambda)^2$.
\begin{equation}
    \mathbb{E}[\|g^k_t\|^2] \leq EL (2+(a-2)\lambda)^2.
\end{equation}
\end{lemma}
\begin{lemma}[\textbf{Bounded Global Gradient}]\label{lemma:2}
For $t\geq 1$, the global gradient has bound as, 
\begin{equation}
    \mathbb{E}[\|f_t\|^2] \leq EL^2(2+(a-2)\lambda)^2 \sum^L_{l=1} \frac{1}{p_{l}^2}.
\end{equation}
%$\mathbb{E}[\|f_t\|^2] \leq EL^2(2+(a-2)\lambda)^2 \sum^L_{l=1} \frac{1}{p_{l}^2}$.
\end{lemma}
\begin{lemma}[\textbf{Bounded Global Gradient Variance}] 
\label{lemma:3}
Under Assumption~\ref{assum:3}, the variance of the global gradient $f_t$ is bounded within $\mathbf{Z}$, which is given as,
\begin{equation}
    \mathbb{E}\|f_t-\bar{f}_t\|^2 \leq L\delta \sum^L_{l=1} \frac{1}{p_{l}^2}.
\end{equation}
%$\mathbb{E}\|f_t-\bar{f}_t\|^2 \leq L\delta \sum^L_{l=1} \frac{1}{p_{l}^2}$.
\end{lemma}
Note that Lemmas~\ref{lemma:2} and~\ref{lemma:3} are different, in the sense that Lemma~\ref{lemma:2} focuses on the actual gradient, whereas Lemma~\ref{lemma:3} is related to data distributions. 
The convergence analysis utilizes Lemmas~1--3 and eSQFL convergence can be proven by \cite{li2019convergence}.
\begin{theorem}[\textbf{eSQFL Convergence}]\label{theorem:1}
Under Assumptions~\ref{assum:1} and~\ref{assum:3} with the learning rate $\eta_t = \frac{2}{\mu t + 2\beta - \mu}$, we obtain
\begin{align}
\mathbb{E}[F(\theta_{t})] - F^{*} 
&\leq \frac{\beta}{\mu}\cdot\frac{\mu \beta \Delta_1 + 2B }{\mu t+ 2\beta-\mu},\label{eq:thm1} \\
\text{where} ~~~~~ \Delta_t &\triangleq \mathbb{E}\|\Theta_t-\Theta^*\|^2, \\
B &= (EL^2(2+(a-2)\lambda)^2 +  L\delta) \sum^L_{l=1} \frac{1}{p_{l}^2}.
\end{align}
%where $\Delta_t \triangleq \mathbb{E}\|\Theta_t-\Theta^*\|^2$, and $B = (EL^2(2+(a-2)\lambda)^2 +  L\delta) \sum^L_{l=1} \frac{1}{p_{l}^2}$.  

Hence, $\lim\limits_{t\rightarrow \infty}\mathbb{E}[F(\theta_{t})] = F^{*}$.
\end{theorem}
Theorem 1 exhibits several insights of eSQFL as follows.
\begin{enumerate}[leftmargin=10pt]
    \item \textit{Failure under extremely poor channels:}
    Consider an extremely poor channel condition, where the server cannot receive $[l, L]$-th model configurations, \textit{i.e.}, $p_{l'}\simeq 0, \forall l' \in [l,L]$. In this case, the RHS of \eqref{eq:thm1} diverges.
    \item \textit{Importance of successful reception:}
    The optimal gap of eSQFL becomes smaller by increasing the communication opportunities. 
    Consider a perfect channel condition, where the RHS of \eqref{eq:thm1} is minimized. 
    By optimizing the SC transmission, the optimality gap is reduced which is referred to {Proposition 1} and {Corollary 1}.
    \item \textit{Other important metrics:} The optimality gap is affected by the local iterations per communication round $E$, balance factor $\lambda$, and the number of layers $L$.
\end{enumerate}
\begin{proposition}[\textbf{Optimal SC Power Allocation}] 
The transmission power allocation $\bm \nu^*$ minimizing the optimality gap 
is given as,
\begin{equation}
    \bm \nu^* = \arg\min_{\bm\nu}\left(\sum^L_{l=1} \mathrm{exp}\left(-\frac{2/\bar{\gamma}}{\nu_l/ u'-\sum^{L}_{l'=l+1}\nu_{l'}}\right)\right) \label{eq:proposition1}
\end{equation}
where $L\geq 2$, $\nu_l > u' \sum^L_{l'=l+1}\nu_{l'}$ for $\forall  l \in [1,L)$, and $\sum^L_{l=1}\nu_l = 1$.
\end{proposition}
\begin{proof}
Substituting the term $p_l$ into Theorem~\ref{theorem:1},  
the optimality gap is minimized by optimizing the power allocation. 
\end{proof}
\begin{corollary}[\textbf{Low SNR, $\bm {L=2}$}] 
For $L=2$, $\bar{\gamma}\to 0$, and $u'\geq (1+\sqrt{5})/2 \approx 1.618$, the optimal power allocation is as, 
\begin{multline}
    (\nu_1^*,\nu_2^*) = \\ \left(-\frac{\sqrt{u'+1}-u'^2+1}{u'^2+u'}, 1+\frac{\sqrt{u'+1}-u'^2+1}{u'^2+u'}\right).
\end{multline}
%$(\nu_1^*,\nu_2^*) =(-\frac{\sqrt{u'+1}-u'^2+1}{u'^2+u'}, 1+\frac{\sqrt{u'+1}-u'^2+1}{u'^2+u'})$.
%( 1-\frac{1-\sqrt{1+u'}}{u'}, \frac{1-\sqrt{1+u'}}{u'})$.
% $\nu_1 \in (0.5,1]$, and $\nu_1 > \max\left\{0.5, {2 u'(1+u')}/\bar{\gamma} \right\}$, the optimality gap is minimized when $\nu_1^* = \frac{u'+\sqrt{1+u'} -1}{u'}$.
\end{corollary}
\begin{proof}
Since $\exp(-x)=1-x$ for $x\to 0$, the RHS of \eqref{eq:proposition1} becomes $2 + \frac{2/\bar{\gamma}}{ \nu_1/u'-(1-\nu_1)}+\frac{2/\bar{\gamma}}{ (1-\nu_1)/u'}$, which is piece-wise convex. Applying the first-order necessary condition (FONC) with respect to $\nu_1$ completes the proof.
% The condition $P_1 > P_2$ is considered. Substitute $L=2$, $\nu_2 = 1 - \nu_1$, and the term $H$ is summarized with $\nu_1$, which is written as 
    % H = \exp\left(-\frac{2/\bar{\gamma}}{\nu_1/u'-(1-\nu_1)}\right)+\exp\left(-\frac{2/\bar{\gamma}}{ (1-\nu_1)/u'}\right).$
% If $\nu_1 > 2 u'(1+u')/{\bar{\gamma}}$, both terms can be approximated in $H$ using the first-order Taylor expansion, yielding, $H \approx 2 + \frac{2/\bar{\gamma}}{ \nu_1/u'-(1-\nu_1)}+\frac{2/\bar{\gamma}}{ (1-\nu_1)/u'}. $
% The approximated $H$ is convex, and the proof is finalized.
 \end{proof}

\begin{table}[t!]    
\small
\caption{List of simulation parameters.}
    \centering
    %\resizebox{\columnwidth}{!}{
    \begin{tabular}{l|r}
    \toprule[1pt]
      \bf{Description}                & \bf{Value}  \\ \midrule
        Number of devices ($N$)       & 10 \\
        Local iterations per communication round ($E$)       & 10 \\
        Epoch ($T$)                   & 100 \\
        Optimizer                     & SGD \\
        Learning rate ($\eta_1$),    & $0.01$\\
        Decaying rate   &  $0.001$\\
        Observable hyperparameter ($a$) & $2$\\
        Number of qubits & $4$\\
        Number of parameters in eSQFL \& Vanilla QFL& $36$\\
%        Number of parameters in {Classical FL} & 56\\
        Number of data per device & $\!128$\\
        Batch size ($D$)   & $32$\\
        \bottomrule[1pt]
    \end{tabular}
    %}
    \label{tab:tab_parameters}
\end{table}
\section{Experiments}\label{sec:6}
%We demonstrate the robustness of eSQFL under various communication environments and various non-IIDness. we consider the following baselines: eSQFL with both CU gate and IPFD, vanilla QFL with QNN, and Classical FL with classic NNs.
%\subsection{Simulation Settings} 
%\subsection{Setup} 

% To evaluate the performance of eSQFL with IPFD and CU gate,we measured the performance of our proposed model, eSQFL, using the top-1 accuracy metric in the MNIST classification task. As quantum machine learning simulation has the computation resource limitation from the lack of input qubits, we adopt mini-MNIST for the classification task, where the dataset is interpolated from region to region into 7x smaller sizes.o verify the performance of eSQFL in various data distribution, we produced dataset with different distributions. Fig.~\ref{Non-IID dist} represents three different mini-MNIST dataset depending on $\alpha$. As $\alpha$ decreases, the data distribution of each local model becomes non-IID.  

\subsection{Experimental Design}\label{sec:6-1}
To corroborate the main analysis and hypothesis of this paper, the experiments are designed as follows:
\begin{itemize}[leftmargin=10pt]
    \item  From Sec.~\ref{sec:5-1}, the derived convergence bound is highly affected by the decoding success probability and non-IIDness. To corroborate these results numerically, we compare the top-1 accuracy of eSQFL in various channel conditions and degrees of non-IIDness with Vanilla QFL (referred to Fig.~\ref{fig:abstract}(a)).
    \item  We investigate the advantage of CU gates that compose eSQNN by designing an experiment which measures entanglement entropy and top-1 accuracy of eSQNN and standard QNNs under the same conditions. Then, the two metrics are compared to demonstrate the advantage of CU gates. 
    \item The increased effectiveness of local training with IPFD compared to IPKD is proven. IPFD trains the local models by regularizing the fidelity of two quantum states. In contrast, IPKD trains local models by ensuring that the small model follows the large model via its prediction. The benchmark scheme comparing the fidelity and top-1 accuracy of IPFD and IPKD is designed. 
    \item According to {Proposition 1} and {Corollary 1}, the convergence bound is minimized by optimal transmission power allocation. To corroborate this, we compare the optimal power allocation scheme to its random power allocation counterpart.
    \item Finally, we conduct experiments by controlling various variables and assess their various impact on the performance. 
\end{itemize}
\begin{figure}[t!]\centering
\begin{tabular}{@{}c@{}c@{}c@{}}
\includegraphics[width=.3333\columnwidth]{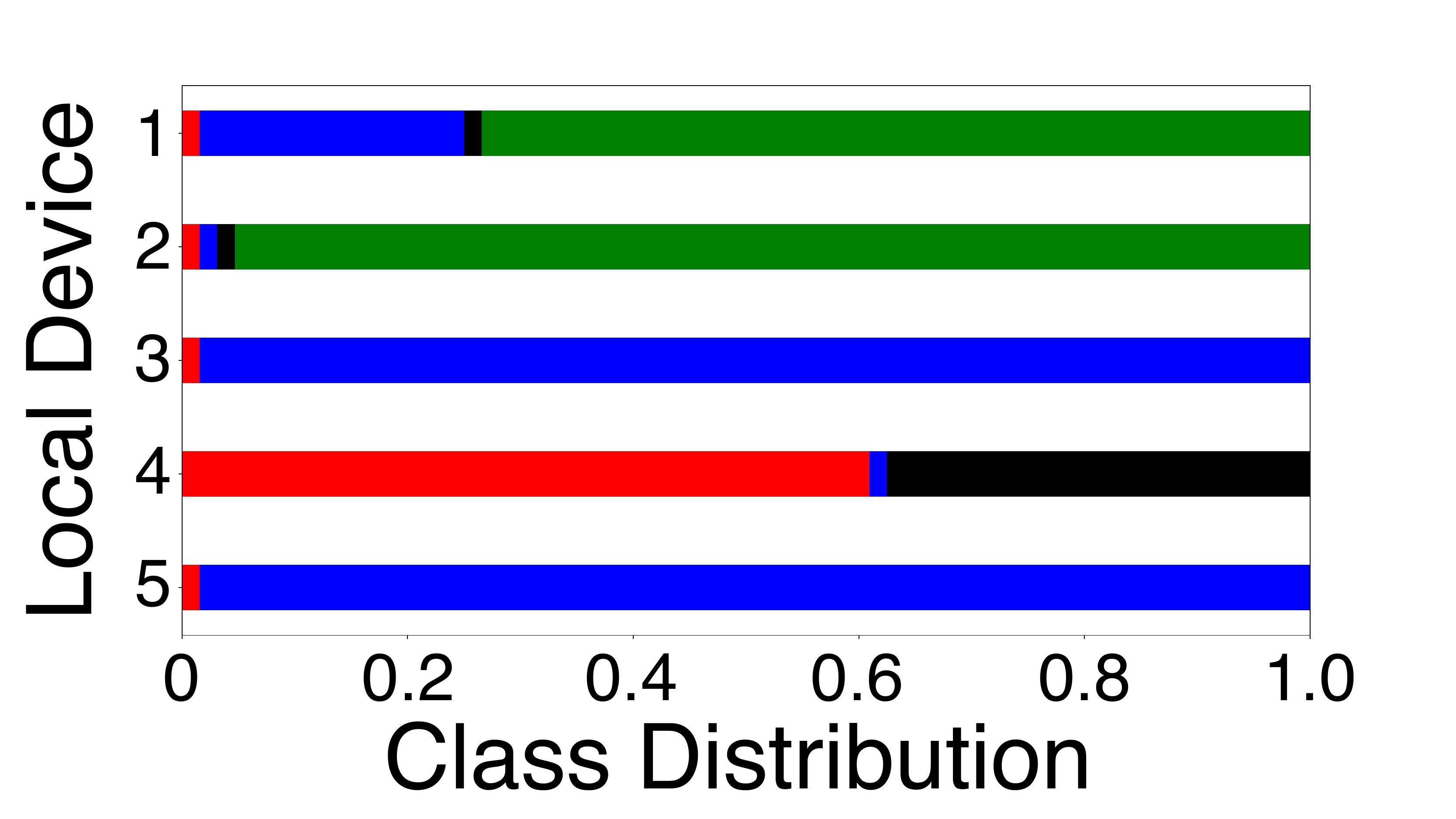} &
\includegraphics[width=.3333\columnwidth]{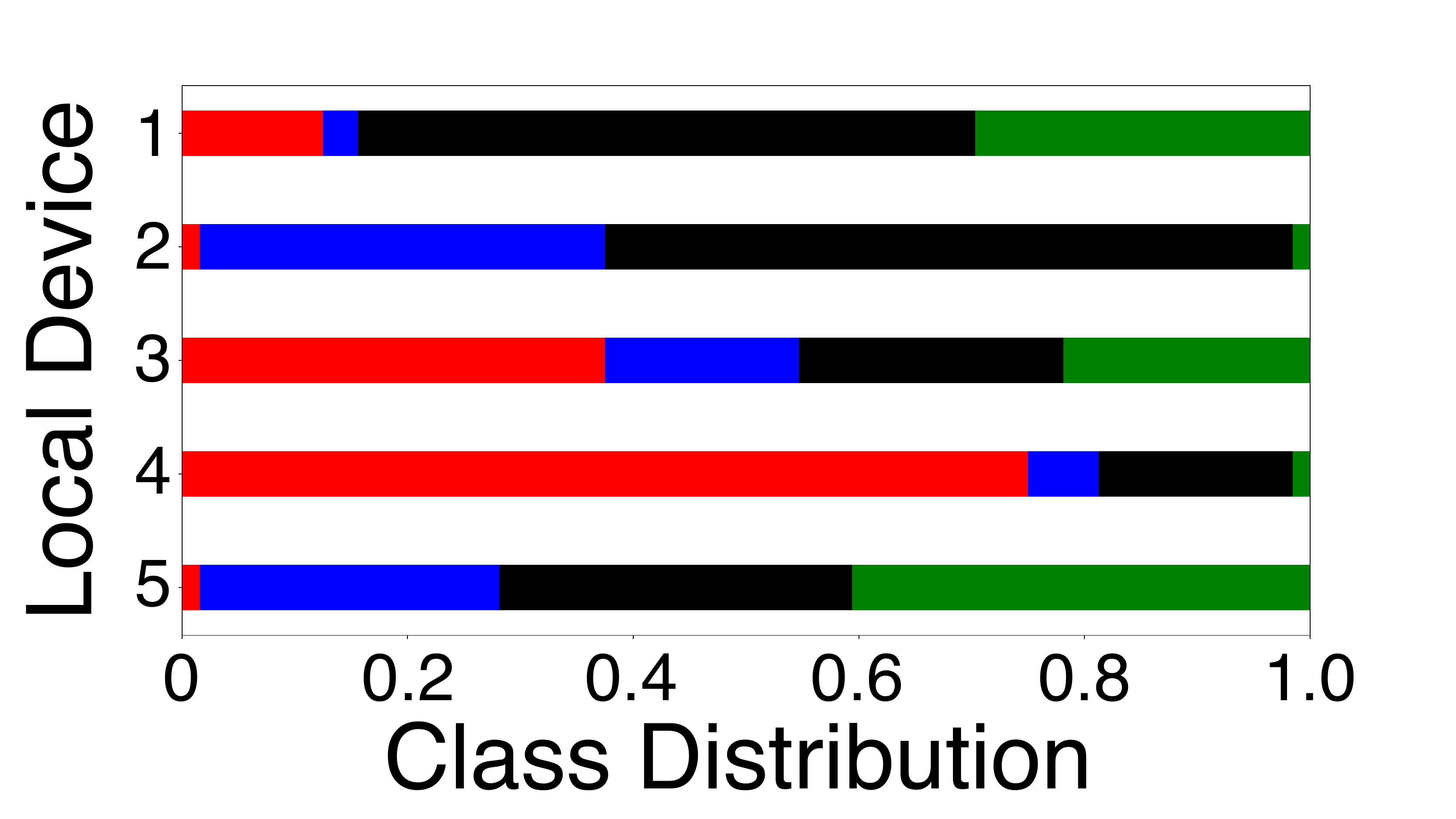} &
\includegraphics[width=.3333\columnwidth]{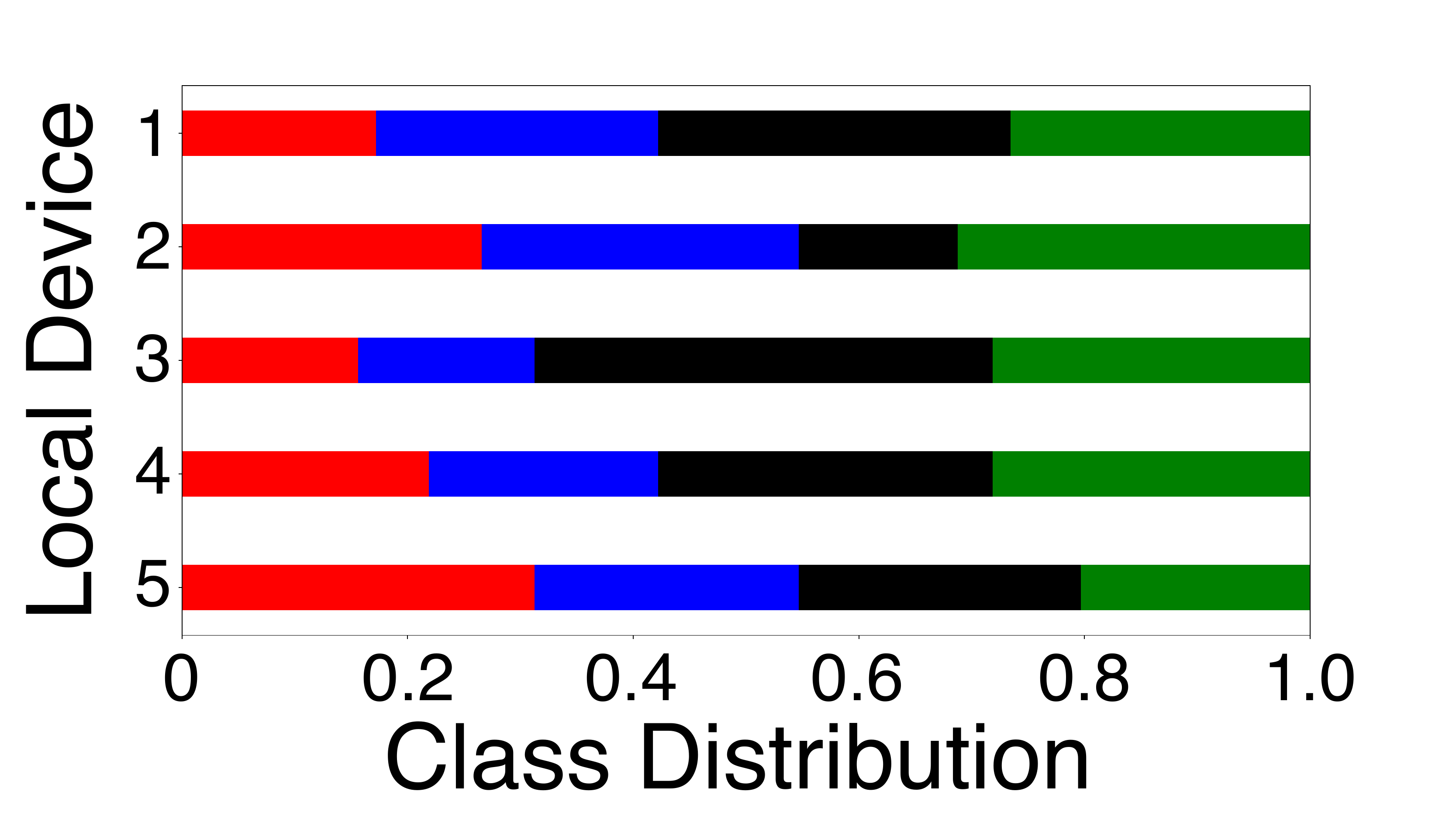} \\
\small (a) $\alpha = 0.1$. & (b) $\alpha = 1.0$. & (c) $\alpha = 10$.
\end{tabular}
%\vspace{-3.5mm}
\caption{Class distributions with different Dirichlet concentration $\alpha$.}\label{Non-IID dist}
%\vspace{-3.5mm}
\end{figure}
For the experiment, eSQFL and Vanilla QFL are evaluated. eSQFL is the proposed model which leverages eSQNN. 
This specific QNN consists of three sub-models named `L1', `L2', and `L3'. 
%Additionally, the number of sub-models to be activated in each iteration can be changed easily, allowing the depth of QNN to be flexible. 
In contrast, Vanilla QFL uses a standard QNN which is made up of basic quantum gates~\cite{chen20}, and does not consider SC and SD~\cite{chen2021federated}. 
Despite the difference in structure, both eSQNN and standard QNN use equivalent number of parameters.
Moreover, we conduct ablation studies on our eSQNN by comparing it with Vanilla SQNN, a depth-controllable yet entanglement-fixed QNN.
Since the performance of QFL suffers under a system with a large number of qubits, many QFL works use a simple dataset~\cite{chen2021federated}. 
In this paper, the MNIST dataset is transformed into a simpler form: the dimension of MNIST data is reduced to $4\times4$ by inter-area interpolation, and only four classes are used (\textit{i.e.}, 0, 1, 2 and 3)~\cite{deng2012mnist}. The four classes are represented with red, blue, black, and green respectively. In addition, Dirichlet distribution is used to investigate non-IIDness of data~\cite{Tzu2019measuring}. Fig.~\ref{Non-IID dist} shows the data distribution with the different values of the Dirichlet concentration ratio $\alpha$. Data with high Dirichlet concentration ratio (\textit{i.e.,} $\alpha=10$) is IID while data with low Dirichlet concentration ratio (\textit{i.e.,} $\alpha=0.1$) is non-IID. 

%
%Finally, classic FL refers to the model utilizing classic NN and FedAvg technique.
To compare IPFD and IPKD, we initialize the parameter of eSQNN identically. The simulation parameters used in these numerical experiments are summarized in Tab.~\ref{tab:tab_parameters}.

\begin{figure}[t!]
\centering
\begin{tabular}{@{}c@{}c@{}c@{}}
\includegraphics[width=.3333\columnwidth]{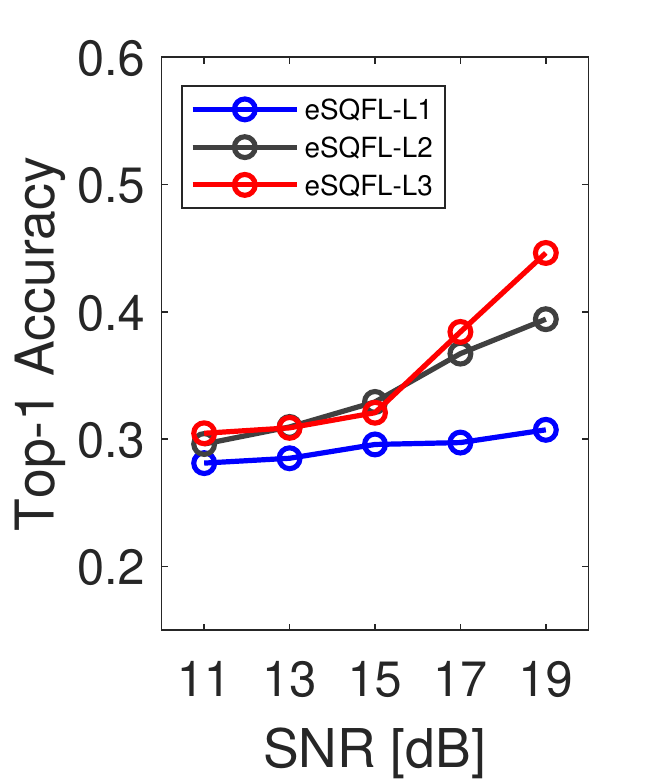}&
\includegraphics[width=.3333\columnwidth]{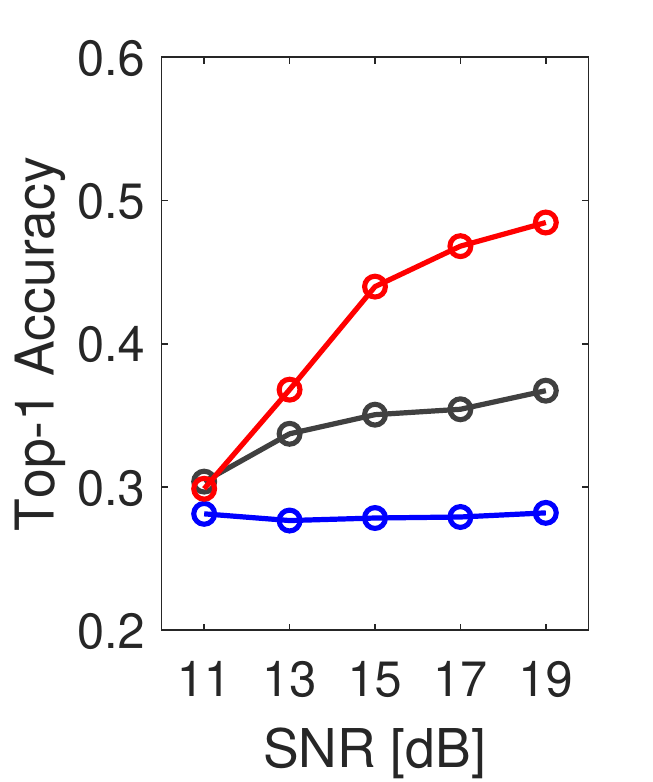}&
\includegraphics[width=.3333\columnwidth]{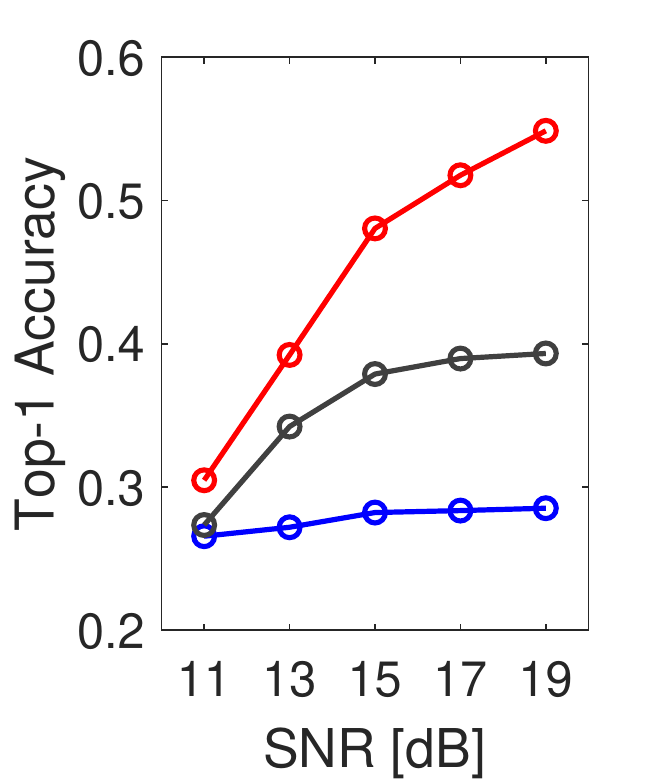}\\[0pt]
\small (a) $\alpha = 0.1$. & (b) $\alpha = 1$. & \small (c) $\alpha = 10$.\\
\end{tabular}
%\vspace{-3.5mm}
\caption{Comparison of top-1 accuracy under various avg. SNR [dB] and $\alpha$.}\label{SNR}
%\vspace{-3.5mm}
\end{figure}
\begin{figure}[t!]
\begin{tabular}{@{}c@{}c@{}c@{}}
\includegraphics[width=.3333\columnwidth]{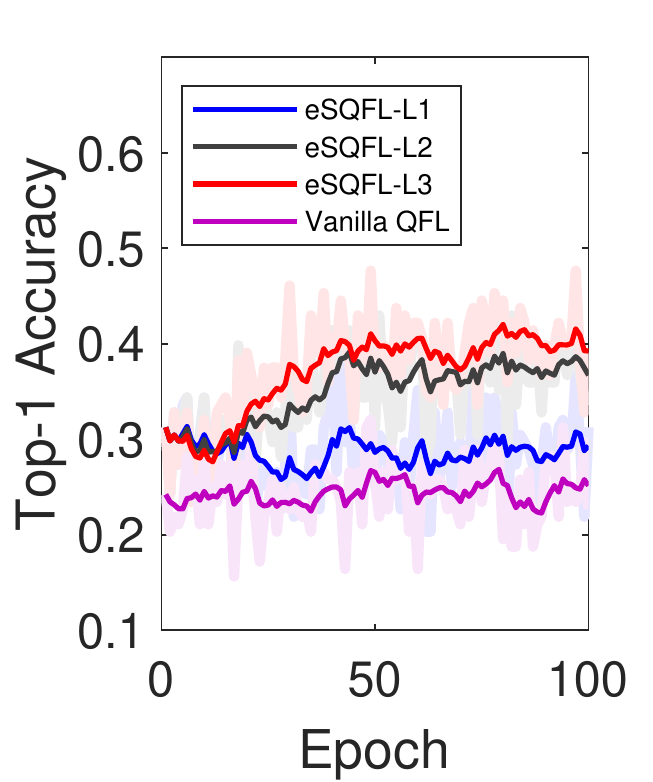}&
\includegraphics[width=.3333\columnwidth]{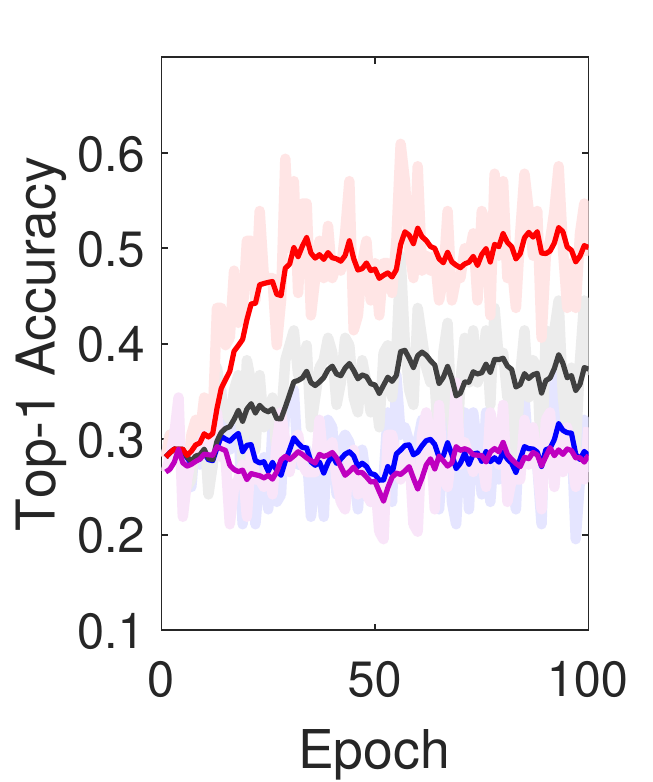}&
\includegraphics[width=.3333\columnwidth]{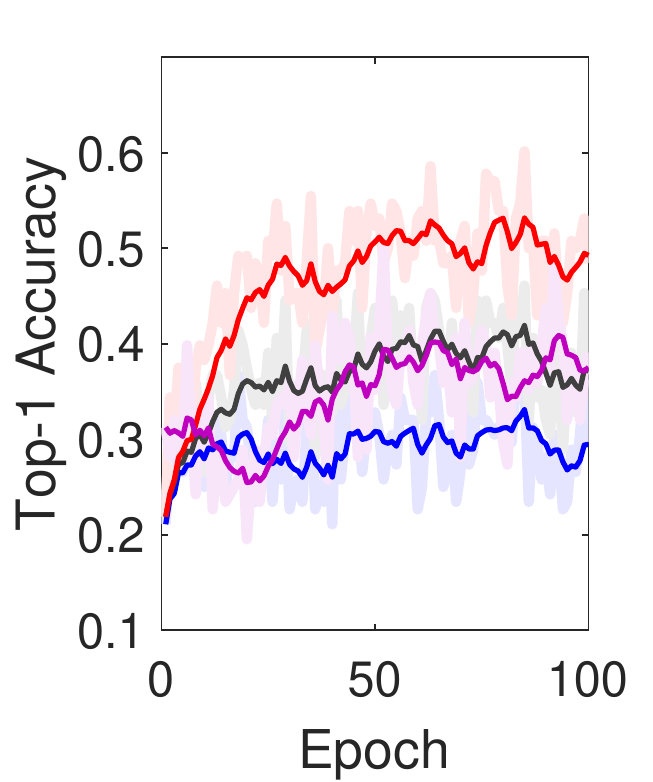}\\
\small (a) $\alpha = 0.1$. &\small (b) $\alpha = 1$. & \small (c) $\alpha = 10$.\\
\end{tabular}
%\vspace{-3.5mm}
\caption{Comparison of top-1 accuracy under various $\alpha~(\bar{\gamma} = 17\,\text{dB})$.}\label{Main_result}
\end{figure}
\subsection{Numerical Results}\label{sec:6-2}
\BfPara{Numerical Results and Convergence Analysis} According to Theorem \ref{theorem:1}, the convergence bound decreases if the decoding success probability increases. 
Fig.~\ref{SNR} shows the performance of eSQFL under various channel conditions obtained through various $\sigma^2$. As $\bar{\gamma}$ increases from $11\,\text{dB}$ to $19\,\text{dB}$, the decoding success probability and top-1 accuracy of the eSQFL with all layers increase. %In addition, the effectiveness of IPFD can be seen in Fig.~\ref{SNR} (c). 
The small models, \textit{i.e.}, eSQFL-L2 and eSQFL-L1, also show improvement in performance along with eSQFL-L3. Especially, eSQFL-L2 shows significant improvement in top-1 accuracy from $28\%$ to $39\%$.
Fig.~\ref{Main_result} shows the top-1 accuracy and convergence of eSQFL and comparison models. When $\bar{\gamma} = 17\,\text{dB}$, the sub-models in eSQFL (\textit{i.e.,} eSQFL-L2, eSQFL-L3) achieve higher accuracy than Vanilla QFL. The final standard deviations of eSQFL under $\bar{\gamma} = 17\,\text{dB}$ are 0.041, 0.051, and 0.066 for eSQFL-L1, eSQFL-L2, and eSQFL-L3, respectively. 

According to {Theorem \ref{theorem:1}}, the data distribution affects the convergence bound of eSQFL. With non-IID data, the convergence bound is widened. As shown in Fig.~\ref{Main_result}, we test various Dirichlet concentration, \textit{i.e.}, $\alpha=\{0.1,1,10\}$. The overall performance of all comparison models decreases as $\alpha$ decreases. 
However, eSQFL shows robustness under non-IID data distribution. Vanilla QFL shows low top-1 accuracy under $\alpha=1$ and $\alpha=0.1$. 
In contrast, eSQFL maintains the top-1 accuracy of $52\%$ and $41\%$ under $\alpha=1.0$ and $\alpha=0.1$ respectively. From the results in Fig.~\ref{SNR} and Fig.~\ref{Main_result}, eSQFL is robust under various channel conditions and non-IID data distribution.

\begin{figure}[t] 
\centering
\begin{tabular}{@{}c@{}c@{}c@{}}
    \multicolumn{3}{c}{\includegraphics[width=0.85\columnwidth]{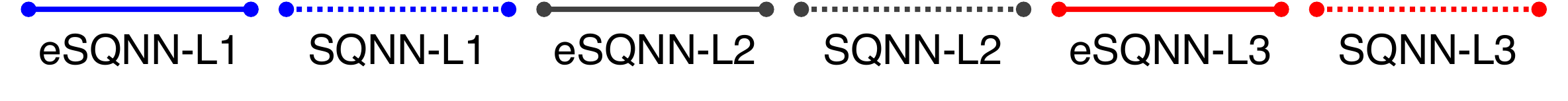}}\\
    \includegraphics[width=.3333\columnwidth]{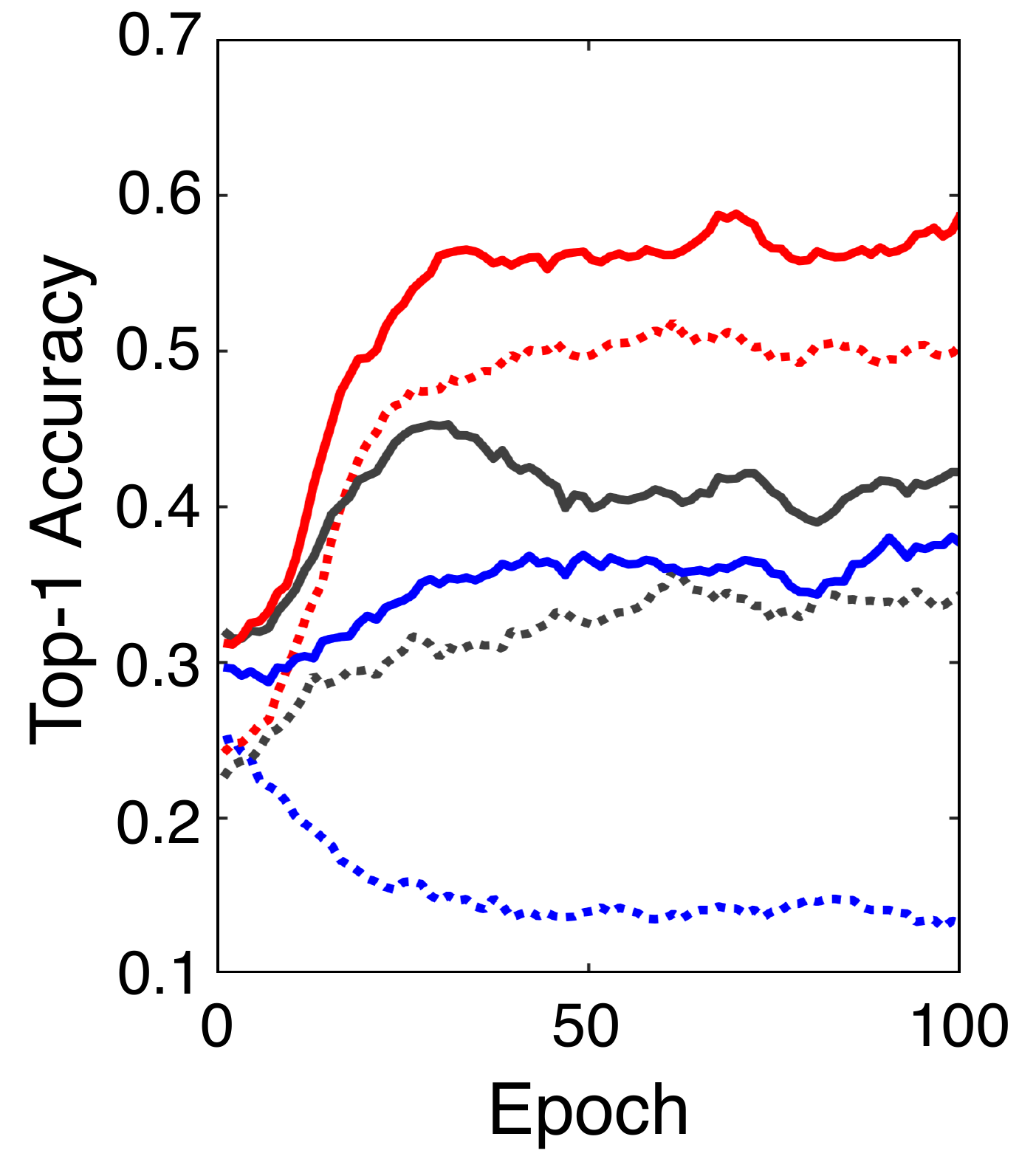}&
    \includegraphics[width=.3333\columnwidth]{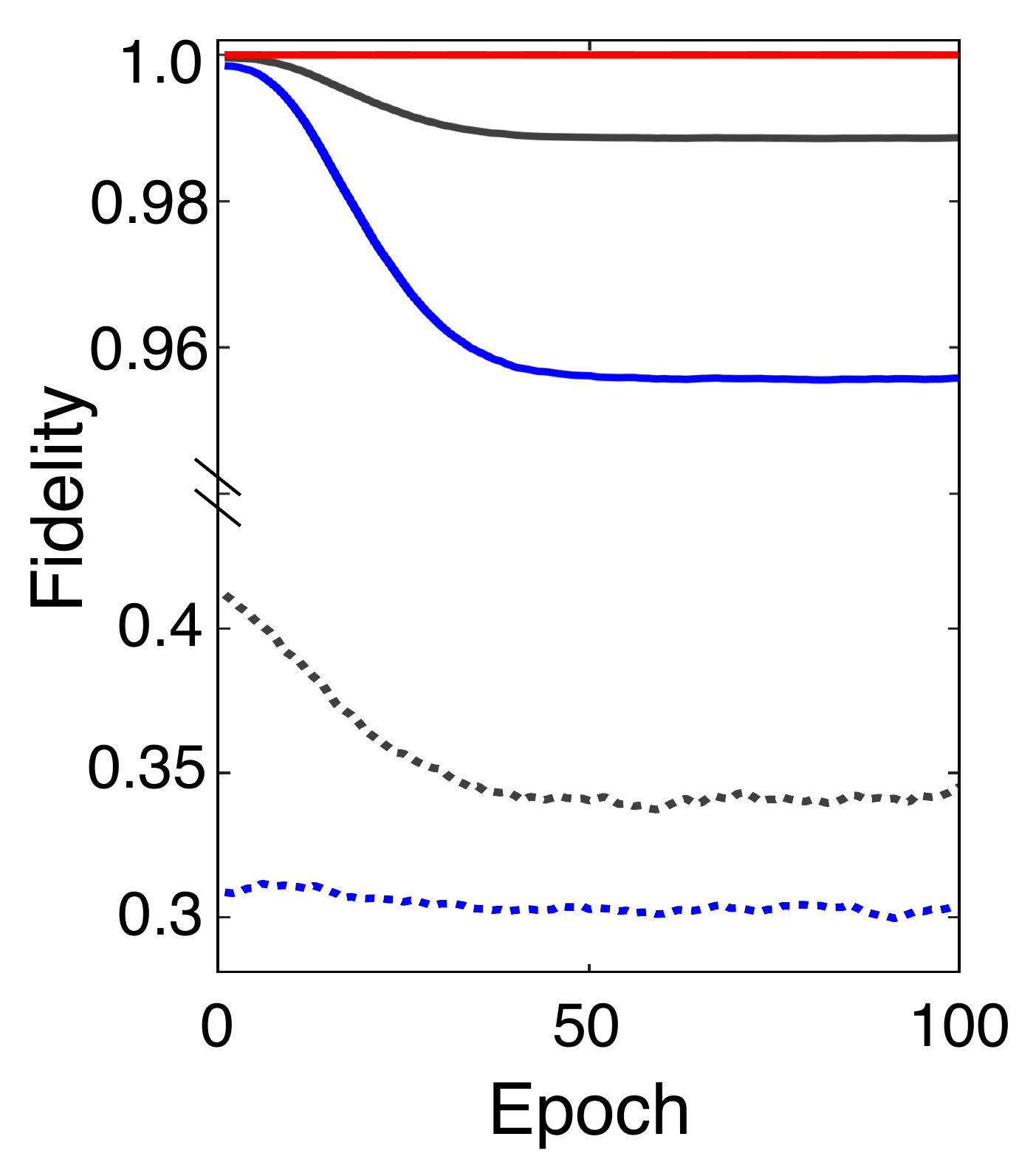}
    & \includegraphics[width=.3333\columnwidth]{Figures/Figure6b.pdf}\\
\small  (a) Top-1 accuracy. & \small  (b) Fidelity. & \small  (c) Entropy.
\end{tabular}
\caption{Model architectural difference (eSQNN vs. Vanilla SQNN).}
\label{fig:eSQNN-vs-SQNN}
\end{figure}
\begin{figure}[t] 
\centering
    \begin{tabular}{@{}c@{}c@{}}
    \includegraphics[width=0.5\columnwidth]{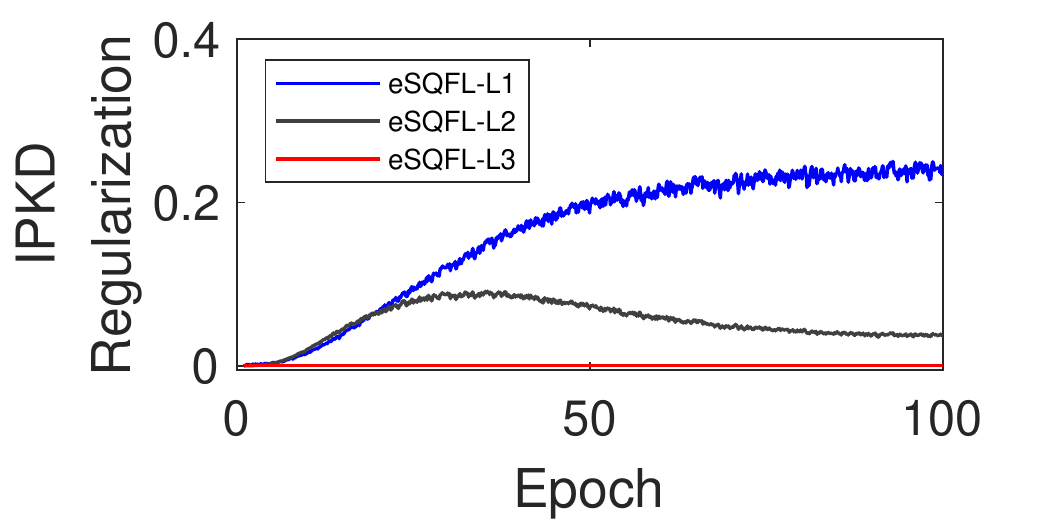} & \includegraphics[width=0.5\columnwidth]{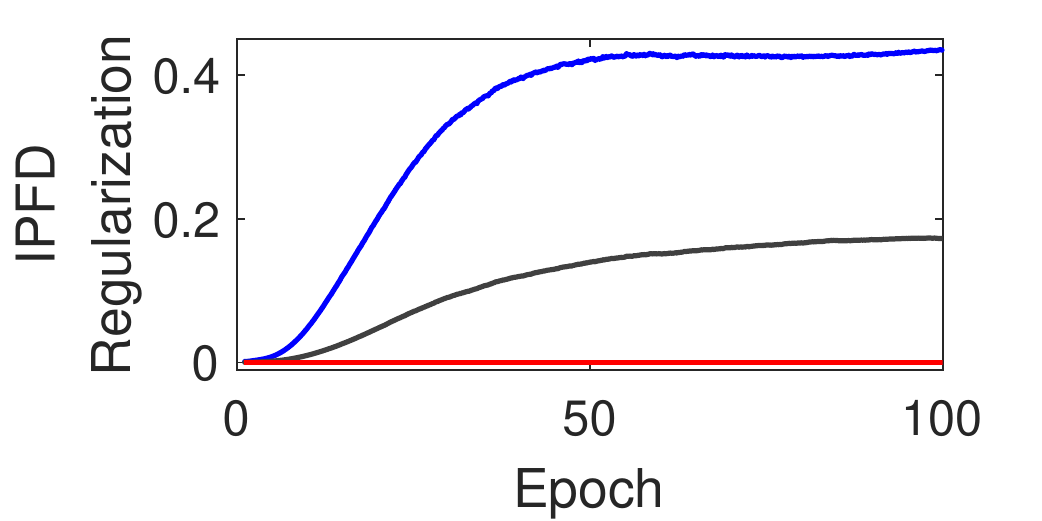}\\
    \small (a) IPKD Regularization. & \small (b) IPFD Regularization.\\
    \end{tabular}
\caption{Comparison of IPFD training algorithm under non-IID and IID.}
\label{fig:IPKDIPFD}
\end{figure}
\begin{figure}[t!] 
\centering
    \begin{tabular}{@{}c@{}c@{}}
        \includegraphics[width=0.5\columnwidth]{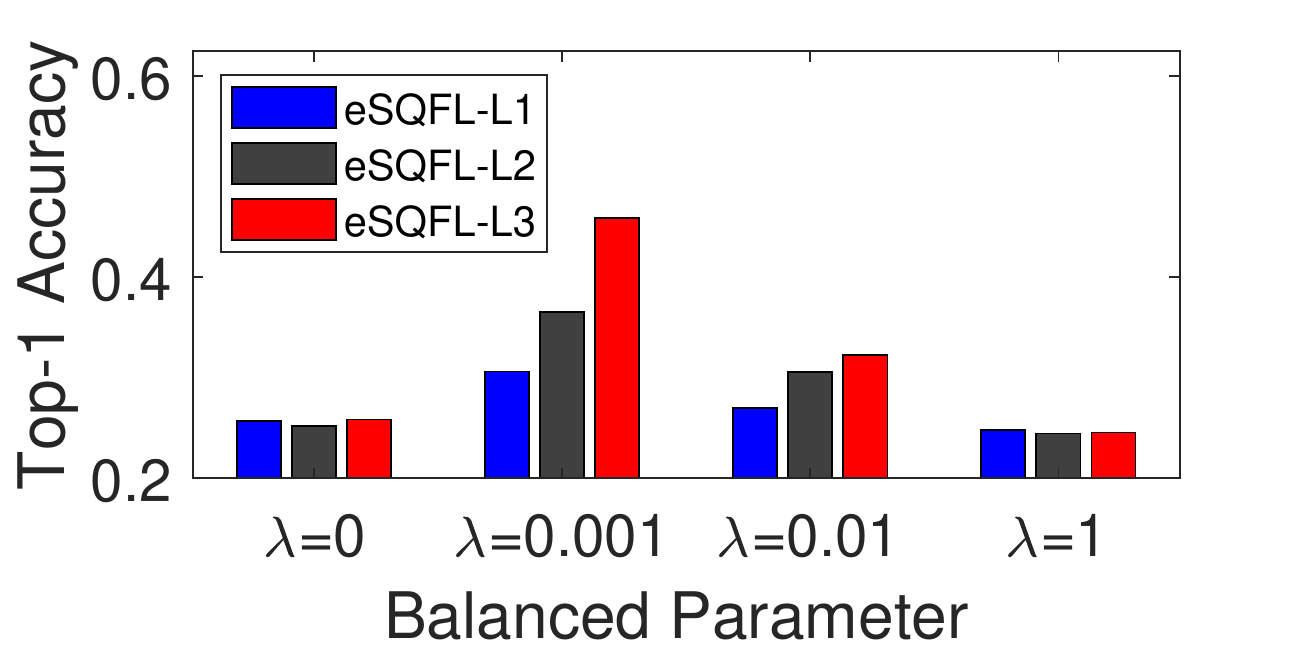} & \noindent 
        \includegraphics[width=0.5\columnwidth]{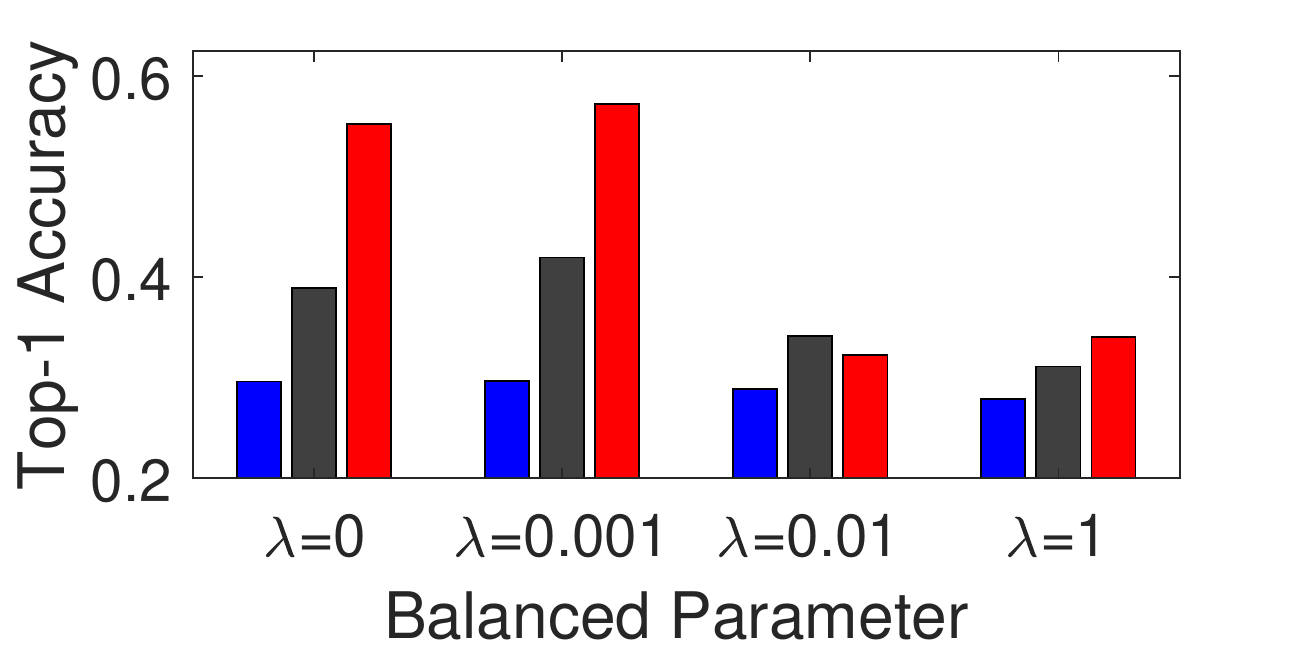}\\
        \small (a) $\alpha=0.1$. &
        \small (b) $\alpha=10$.
   \end{tabular}
%\vspace{-3mm}
\caption{Comparison of fidelity training algorithm under non-IID and IID.}
%\vspace{-1mm}
\label{fig:fidel_bar}
\end{figure}

\BfPara{Structural Advantage of eSQNN}
In this subsection, we investigate the general performance, fidelity, and entanglement entropy of eSQNN. We conduct ablation studies corresponding to model architecture (\textit{i.e.}, eSQNN and Vanilla SQNN). 
Fig.~\ref{fig:eSQNN-vs-SQNN} shows the experiment results. As shown in Fig.~\ref{fig:eSQNN-vs-SQNN}(a), eSQNN shows better top-1 accuracy than Vanilla eSQNN. Especially, eSQNN-L3 achieves a $20\%$ performance improvement over Vanilla SQNN. 
When eSQNN is used, the quantum state (\textit{i.e.}, knowledge) is successfully distilled to the small model as shown in Fig.~\ref{fig:eSQNN-vs-SQNN}(b). In contrast, Vanilla  SQNN fails to distill the knowledge to its sub-model.
To understand why its model is successfully trained, we calculate the entanglement entropy (referred to Sec.~\ref{sec:3}). 
Fig.~\ref{fig:eSQNN-vs-SQNN}(c) exhibits the von Neumann entanglement entropy of each layer of eSQNN and Vanilla SQNN. The entanglement entropy of eSQNN is less than Vanilla SQNN for all layers. It means that the event of exceeding the entropy threshold aforementioned in Sec.~\ref{sec:6}, \textit{i.e.,} $\mathbbm{1}_{\text{train}} = 0$, rarely occurs compared to Vanilla SQNN. This underscores that eSQNN is more robust to barren plateaus than Vanilla SQNN.

\BfPara{Effectiveness of IPFD}
To investigate the effectiveness of IPFD used in eSQNN local training, we compare the results of local training using IPFD regularizer to IPKD regularizer. 
Fig.~\ref{fig:IPKDIPFD} (a)/(b) show the learning curve of $\lcal_{FD}$ and $\lcal_{KL}$, respectively.
The learning curve of IPFD starts at $\lcal_{FD}= 0$ due to the fidelity $\fcal(\psib_l, \psib_L) \approx 1$. As eSQNN is trained, the fidelity decreases and converges to 0.955 for L1 and 0.987 for L2.
In the learning curve of $\lcal_{KD}$, the curve has a tendency to decrease and converge. 
However, the fluctuation of IPKD regularization is larger than IPFD, especially in eSQFL-L1. 
This is because the KL divergence becomes unstable when the difference between the two distributions is large, \textit{i.e.}, the overlapping area between the distributions is small. Then, if there is no overlapping area, it diverges.
In contrast, the aforementioned phenomena does not occur in IPFD regularization because IPFD regularizer is bounded from 0 to 1.
Therefore, IPFD regularization provides more stable noise to its eSQNN than IPKD regularization.

{\color{red}{
\begin{table}[t!]
\small
\centering
\caption{Top-1 accuracy comparison with ($\bm \nu^*$) and without optimization. }
\begin{tabular}{c||ccc|cc}
    \toprule[1pt]   
     &  \multicolumn{3}{c|}{\textbf{$L=3$}} &  \multicolumn{2}{c}{\textbf{$L=2$}}\\
    \textbf{Condition} & $l=1$ & $l=2$ & $l=3$ & $l=1$ & $l=2$ \\\midrule
    with optimization ($\bm \nu^*$)  & $\mathbf{29.6}$& $\mathbf{40.1}$& $\mathbf{55.8}$& $\mathbf{33.4}$& $\mathbf{50.7}$ \\
    w.o. optimization & $29.5$& $39.3$& $50.7$& $33.0$& $48.1$ \\
\bottomrule[1pt]
\end{tabular}
\label{tab:proof}
% \vspace{-5mm}
\end{table}}}
  
\BfPara{Impact on Optimal Power Allocation}
We verify the proofs of {Proposition 1}, and {Corollary 1}. When $L=3$, we calculate the power allocation variable as $\bm \nu^*=\{0.8909, 0.0989, 0.0102\}$ by non-convex optimization. When $L=2$, we obtain $\bm \nu^*=\{0.8969, 0.1059\}$, where the comparison of power allocation is set to $\bm \nu = \{0.9170, 0.0820, 0.001\}$ for $L=3$, and $\bm \nu = \{0.8333, 0.1667\}$ for $L=2$. 
The final accuracy is Tab.~\ref{tab:proof}. Compared to the eSQFL with $\bm \nu$, the eSQFL with $\bm \nu^*$ achieves $10.1\%$ higher top-1 accuracy when $L=3$ and $5.41\%$ higher top-1 accuracy when $L=2$. Thus, we corroborate that the optimal power allocation minimizes the convergence bound.

\BfPara{Impact on Balanced Parameter}
The balanced parameter $\lambda$ is an important parameter in eSQNN local training. Fig.~\ref{fig:fidel_bar} shows the top-1 accuracy according to $\lambda$ in various data distributions (\textit{i.e.}, $\alpha=0.1$ and $\alpha=10$). With finitely adjusted IPFD parameter ($\lambda^*=0.01$), eSQNN shows the highest top-1 accuracy under non-IID data distribution (\textit{i.e.}, $\alpha=0.1$). In addition, by not using IPFD ($\lambda=0$) or only using IPFD ($\lambda=1$), eSQNN fails to classify the mini-MNIST dataset. Under IID data distribution (\textit{i.e.}, $\alpha=10$), eSQNN with $\lambda^*=0.01$ outperforms eSQNN with only using label training ($\lambda=0$) about $1.03\%$. From the result, we recommend utilizing eSQNN with $\lambda^*=0.001$ for robust performance in both IID and non-IID data distribution.
\iffalse
{\color{blue}{ 
    \BfPara{Other Important Metrics} 
    We investigate the impact on important metrics (\textit{e.g.}, the number of local devices, the number of local iterations per communication round, and the number of layers in eSQNN).
    As shown in Fig.~[REF](a), the top-1 accuracy of eSQFL increases as the number of local devices increases. When the number of federating devices are \tred{X}, the top-1 accuracy is highest, \textit{i.e.,} \tred{XX.x\%} for eSQFL-L1,  \tred{YY.y\%} for eSQFL-L2, and  \tred{ZZ.z\%} for eSQFL-L3, respectively. 
    In addition, the number of local iterations also affects the optimality gap. Fig.~[REF](b) shows the top-1 accuracy of eSQFL corresponding to the number of local iterations per communication round. When \tred{$ E = X$}, eSQFL achieves the highest top-1 accuracy, \textit{i.e.},  \tred{XX.x\%} for eSQFL-L1,  \tred{YY.y\%} for eSQFL-L2, and  \tred{ZZ.z\%} for eSQFL-L3, respectively. 
    As shown in Tab.~\ref{tab:proof}, eSQNN-L3 shows the highest top-1 accuracy. The second highest top-1 accuracy returns to eSQFL-L2 under $L=2$. This is because the optimal gap decreases as $L$ decreases (referred to {Theorem \ref{theorem:1}}).
}}
\fi
\section{Conclusions}\label{sec:7}
In this paper, we developed a depth-adjustable QNN architecture, and proposed a novel QFL framework based on wireless communications, termed eQSNN and eSQFL, respectively.
% We faced various challenges while constructing a dynamic QNN, \textit{i.e.}, there were several weak problems with quantum circuits. 
To control the level of entanglement and reduce its entropy, we applied CU gates to the eSQNN architecture. To mitigate the inter-depth interference inspired from the fidelity in quantum information theory, we introduced a novel IPFD regularizer. Finally, to cope with various channel conditions, we applied SC across multiple depths and optimized the SC power allocation by deriving and minimizing the convergence bound of eSQFL.
%이 작업을 통해 변하는 통신 환경과 NISQ 제한이 존재함에도 불구하고 안정적인 결과를 보이는 양자 연합학습 모델을 제시하였다.
%그 외에도, fidelity regularizer라는 양자 컴퓨팅과 적합한 최적화 방법을 제시하여, 고전 방법들에 의존하지 않고 양자 기계학습의 에러율을 낮추기도 하였다.
%이 논문에서 모델의 강점들을 이론적으로 확인하였으니, 다음으로는 실질적 효용성을 검토하고자 한다. 후속 작업에서는 이 모델을 다양한 제한점을 가진 여러 실제 상황에 적용해볼 것이다.  
In conclusion, we were able to propose a QFL model that shows stable performance despite the NISQ limitation and variable channel conditions.
Additionally, the fidelity regularizer was also designed. This novel method decreases the error rate of QML in a way that is exclusively suitable with QC, instead of depending on classical optimization methods. 
Since the strengths of our model has been theoretically corroborated in this paper, we will go on to test the realistic efficacy of the model. In our future research directions, we will apply it to a plethora of real-life scenarios with various limitations.

% To resolve these problems, we propose an eSQNN that can control entanglement during the local training phase and can be used with a multi-depth configuration. 
% In addition, we propose IPFD to increase the efficiency of training eSQNN, and eSQFL is also configured by applying eSQNN with SFL which utilizes SC and SD in FL.
% We first analyze the convergence to the best of our knowledge using quantum properties in QFL and dynamic QFL.
% Through the theoretical convergence bound, we demonstrate that communication has a very large effect on convergence.
% From the extensive experiments, it was confirmed that all the techniques proposed in this paper outperform the existing techniques.
% Finally, the method proposed in this paper is a structure that can be used in both real quantum devices and CPU devices. If eSQFL is used in a quantum device, its complexity is $\ocal(n)$ where $n$ is the number of parameters.

\appendix
\begin{table}[t!]    
\caption{List of notations.}
\small
    \centering
    \begin{tabular}{c|l}
    \toprule[1pt]
      \bf{Notation} & \bf{Description} \\ \midrule
        $K$ & Number of local devices, $[1,\cdots,k,\cdots,K]$\\
        $L$ & Number of local eSQNN blocks, $[1,\cdots,l,\cdots,L]$ \\
        $E$ & Number of local iterations, $[1,\cdots,e,\cdots,E]$\\
        $T$ & Number of communication rounds, $[1,\cdots,t,\cdots,T]$\\
        $\bm \Xi$ & Binary mask, $\bm \Xi = \{\Xi_1,\cdots,\Xi_l,\cdots,\Xi_L\}$ \\
        $\psib$ & Quantum state \\
        $\rho$ & Reduced density matrix \\ 
        $S_l(\rho)$ & Entanglement entropy of $\rho$ over subsystem $l$\\
        $\bm Z$   & Whole data, $\bm Z = \{\zeta^1,\cdots,\zeta^k,\cdots,\zeta^K\}$ \\
        $\bm \nu$ & Power allocation for SC, $\bm \nu = \{\nu_1,\cdots,\nu_l,\cdots,\nu_L\}$ \\
        $\alpha$ &   Dirichlet concentration \\
        \bottomrule[1pt]
    \end{tabular}
    \label{tab:notations}
% \vspace{-3mm}
\end{table}

\subsection{Parameter Shift Rule}\label{sec:parameter-shift}
Parameter shift rule~\cite{mitarai18}, one of the most known quantum gradient calculators, is utilized to train the model. Subsequently, the eSQNN is trained accordingly using the zeroth stochastic gradient descent algorithm, \textit{e.g.}, quantum natural gradient. 
Consider that eSQNN consists of $I$ trainable parameters, \textit{i.e.}, $\thetab^k_{t,e} = [\theta^k_{t,e,1}, \cdots, \theta^k_{t,e,i}, \cdots, \theta^k_{t,e,I}]$. 
Then, the partial derivative of $k$-th device's $c$-th observable over parameter $\theta^k_{t,e,i}$ is given as follows,
\begin{equation}\label{eq:psr-1}
    \frac{\partial \langle V_c \rangle_{\thetab^k_{t,e}}}{\partial \theta^k_{t,e,i}} = 
    \frac{\langle V_c \rangle_{\thetab^k_{t,e} + \varepsilon\mathbf{e}_i} - \langle V_c \rangle_{\thetab^k_{t,e} - \varepsilon\mathbf{e}_i} }{2\varepsilon}
\end{equation}  
where $\mathbf{e}_i$ denotes the $i$-th standard basis, and $\varepsilon\in (0, \pi/2]$. 
We calculate the loss gradient using \eqref{eq:psr-1}. % We will elaborate on the processes in Appendix~\ref{sec:proof-lemma1}.
\subsection{Proof of Lemma 1}\label{sec:proof-lemma1}
The true label is class $c$ and the predictions and its derivative is canceled out due to the definition of cross-entropy. Then, the cross-entropy loss is simplified as follows,
%$\lcal_{CE} = - \log  p(y^{k,l,c}_{t,e}|\bfx)$.
\begin{equation}
    \lcal_{CE} = - \log  p(y^{k,l,c}_{t,e}|\bfx).
\end{equation}

Hereafter, we denote $\hat{y}_c = y^{k,l,c}_{t,e}$, and $ p(\hat{y}_c) = p(y^{k,l,c}_{t,e}|\bfx)$. Let's denote the partial derivative of cross-entropy loss and fidelity loss as, 
% $G_1 = \frac{\partial\lcal_{CE}}{\partial \theta^k_{t,e,i}} = \frac{1}{p(\hat{y}_c)}\cdot \frac{\partial p(\hat{y}_c)}{\partial \theta^k_{t,e,i}}$ and $G_2  = \frac{\partial \fcal (\psib^{k,L}_{t,e,\bfx}, \psib^{k,l}_{t,e,\bfx})}{\partial \theta^{k}_{t,e,i}}$.
\begin{align}
G_1 &= \frac{\partial\lcal_{CE}}{\partial \theta^k_{t,e,i}} = \frac{1}{p(\hat{y}_c)}\cdot \frac{\partial p(\hat{y}_c)}{\partial \theta^k_{t,e,i}}, \\
G_2  &= \frac{\partial \fcal (\psib^{k,L}_{t,e,\bfx}, \psib^{k,l}_{t,e,\bfx})}{\partial \theta^{k}_{t,e,i}}.
\end{align}

By the triangle inequality, the partial derivative of \eqref{eq:loss} is bounded as follows,
\begin{equation}
    \Big|\frac{\partial \lcal^{k,l}_{t,e}}{\partial \theta^k_{t,e,i}} \Big|  \leq 
\sum_{(\bfx,\bfy) \in \zeta^k} \Big[\frac{\lambda}{D} |G_1| + \frac{1-\lambda}{D}|G_2|\Big].
\label{lem2start}
\end{equation}

We have 
% $G_1 = a \Big(\sum\limits^{C}_{c' \geq 1, c' \neq c} \hat{y}_{c'}/\sum\limits^{C}_{c=1} \hat{y}_c \Big)\cdot\frac{\partial \langle V_{c}\rangle_{\thetab^{k}_{t,e}\odot \sum^{l}_{l'=1}\Xi_{l'}}}{\partial \theta^{k}_{t,e,i}}$.
\begin{equation}
    G_1 = a \left(\sum\limits^{C}_{c' \geq 1, c' \neq c} \frac{\hat{y}_{c'}}{\sum\limits^{C}_{c=1} \hat{y}_c }\right)\cdot\frac{\partial \langle V_{c}\rangle_{\thetab^{k}_{t,e}\odot \sum^{l}_{l'=1}\Xi_{l'}}}{\partial \theta^{k}_{t,e,i}}.
\end{equation}

The bound of $G_1$ obtained as follows,
%$|G_1| \leq a \Big| \frac{\partial \langle V_{c'}\rangle_{\thetab^k\odot \Xi_{l}}}{\partial \theta^k_{t,e,i}} \Big| \leq a$.
\begin{equation}
    |G_1| \leq a \Big| \frac{\partial \langle V_{c'}\rangle_{\thetab^k\odot \Xi_{l}}}{\partial \theta^k_{t,e,i}} \Big| \leq a.
\end{equation}

The former step is due to $\sum\limits^{C}_{c' \geq 1, c' \neq c} \hat{y}_{c'}\leq \sum\limits^{C}_{c=1} \hat{y}_c$, and the latter step is because the gradient using parameter shift rule is bounded to $1$~\cite{mitarai18}.
The term $G_2$ and its bound are given as,
% $G_2 = 2|\langle\psib^{k,L}_{t,e,\bfx}|\psib^{k,l}_{t,e,\bfx}\rangle|\cdot     \frac{\partial \langle\psib^{k,L}_{t,e,\bfx}|\psib^{k,l}_{t,e,\bfx}\rangle }{\partial \theta^{k}_{t,e,i}}$, and $|G_2|\leq 2\Big|\frac{\partial \langle\psib^{k,L}_{t,e,\bfx}|\psib^{k,l}_{t,e,\bfx}\rangle }{\partial \theta^{k}_{t,e,i}}\Big| \leq 2$. 
\begin{align}
    G_2 &= 2|\langle\psib^{k,L}_{t,e,\bfx}|\psib^{k,l}_{t,e,\bfx}\rangle|\cdot     \frac{\partial \langle\psib^{k,L}_{t,e,\bfx}|\psib^{k,l}_{t,e,\bfx}\rangle }{\partial \theta^{k}_{t,e,i}},\\
    |G_2|&\leq 2\Big|\frac{\partial \langle\psib^{k,L}_{t,e,\bfx}|\psib^{k,l}_{t,e,\bfx}\rangle }{\partial \theta^{k}_{t,e,i}}\Big| \leq 2.
\end{align}

The former step is due to $|\langle\psib^{k,L}_{t,e,\bfx}|\cdot\psib^{k,l}_{t,e,\bfx}\rangle|^2 \leq 1$, and the latter step is due to parameter shift rule.
Substituting the bound of $G_1$ and $G_2$ into LHS of \eqref{lem2start}, we have the bound, 
%$\Big|\frac{\partial \lcal^{k,l}_{t,e}}{\partial \theta^k_i} \Big| \leq 2+(a-2)\lambda$.
\begin{equation}
    \Big|\frac{\partial \lcal^{k,l}_{t,e}}{\partial \theta^k_i} \Big| \leq 2+(a-2)\lambda.
\end{equation}

Calculate LHS of \eqref{lem2start} for all $i \in [1, I]$, the loss gradient is obtained, and its gradient is bounded as, 
%$\|\nabla_{\thetab^k_{t,e}} \lcal^{k,l}_{t,e} \| \leq 2+(a-2)\lambda$.
\begin{equation}
    \|\nabla_{\thetab^k_{t,e}} \lcal^{k,l}_{t,e} \| \leq 2+(a-2)\lambda.
\end{equation}

Applying these results to $g^k_t$, we complete the proof.

\subsection{Proof of Lemma 2}\label{sec:proof-lemma2}
We expand the global gradient $f_t$ as follows,
\begin{align}
    \|f_t\|^2 &= \Big\|\sum^L_{l=1}\frac{1}{Kp_{l}} \sum^K_{k\in |X_l|} g^k_t \odot \Xi_l\Big\|^2 \\
    &\leq \frac{L}{K} \sum^L_{l=1} \frac{1}{p_{l}^2} \sum^K_{k=1} \| g^k_t \odot  \Xi_l\|^2 \\
    &\leq \frac{L}{K} \sum^K_{k=1}\sum^L_{l=1} \frac{1}{p_{l}^2} \cdot  \| g^k_t\|^2 .\label{eq:lem2-solution}
\end{align}

The first step is due to Jensen's inequality, \textit{i.e.},
\begin{equation}
\|\sum^K_{k=1} x_k\|^2 \leq K \sum^K_{k=1} \|x_k\|^2 
\end{equation}
and the next step is due to Cauchy-Schwarz inequality, \textit{i.e.},  $\|X\odot \Xi\|^2 \leq \|X\|^2$~\cite{infocom2022baek}. Combining Lemma \ref{lemma:1} and latter term of \eqref{eq:lem2-solution}, we finalize the proof. 

\subsection{Proof of Lemma 3}\label{sec:proof-lemma3}
\iffalse
According to \eqref{eq:global} and Assumption~\ref{assum:3}, the distance between $f_t$ and $\bar{f}_t$ is as 
\begin{equation}
    \|f_t-\bar{f}_t\|^2 &= \|  \sum^L_{l=1}  \frac{1}{Kp_{l}}  \sum^K_{k\in |X_l|}(g^k_t-\bar{g}^k_t)\odot \Xi_l\|^2 ,\\
    &\leq \frac{L}{K} \sum^L_{l=1} \sum^K_{k=1} \frac{1}{p_{l}^2} \left\|g^k_t-\bar{g}^k_t\right\|^2.
\end{equation}
% $\|f_t-\bar{f}_t\|^2 = \|  \sum^L_{l=1}  \frac{1}{Kp_{l}}  \sum^K_{k\in |X_l|}(g^k_t-\bar{g}^k_t)\odot \Xi_l\|^2 \leq \frac{L}{K} \sum^L_{l=1} \sum^K_{k=1} \frac{1}{p_{l}^2} \left\|g^k_t-\bar{g}^k_t\right\|^2$. 
This step is due to Jensen's inequality.
\fi
According to \eqref{eq:global} and Assumption~\ref{assum:3}, the distance between $f_t$ and $\bar{f}_t$ is as, 
% $\|f_t-\bar{f}_t\|^2 = \Big\|  \sum^L_{l=1}\nolimits  \frac{1}{Kp_{l}}  \sum^K_{k\in |X_l|}\nolimits(g^k_t-\bar{g}^k_t)\odot \Xi_l\Big\|^2\leq \frac{L}{K} \sum^L_{l=1} \sum^K_{k=1} \frac{1}{p_{l}^2} \left\|g^k_t-\bar{g}^k_t\right\|^2$.
\begin{align}\|f_t-\bar{f}_t\|^2 &= \Big\|  \sum^L_{l=1}  \frac{1}{Kp_{l}}  \sum^K_{k\in |X_l|}(g^k_t-\bar{g}^k_t)\odot \Xi_l\Big\|^2 \\
&\leq \frac{L}{K} \sum^L_{l=1} \sum^K_{k=1} \frac{1}{p_{l}^2}. %\left\|g^k_t-\bar{g}^k_t\right\|^2.
\end{align}

This step is due to Jensen's inequality.
With Assumption~\ref{assum:3}, we have $\mathbb{E}\|g^k_t-\bar{g}^k_t\|^2 \leq \sigma_k^2$. Combining these results finalizes the proof.
\subsection{Completing Proof of Theorem 1}\label{sec:proof-theorem1}
Using \eqref{eq:global}, the distance between $\Theta_{t+1}$ to the optimal is as,
\begin{align}
\|\Theta_{t+1} - \Theta^*\|^2 = & \|\Theta_t - \eta_t f_t - \Theta^* +\eta_t \bar{f}_t -\eta_t \bar{f}_t\|^2 \label{eq:lem3-prob}\\
= &\underbrace{\|\Theta_{t} - \eta_t \bar{f}_t - \Theta^*\|^2}_{G_3} \\ & + \underbrace{2\eta_t \langle \Theta_t - \Theta^* - \eta_t f_t, \bar{f}_t - f_t \rangle}_{G_4} \\ & 
+ \underbrace{\eta_t^2 \|f_t - \bar{f}_t\|^2}_{G_5} .\label{eq:lem2-bound} 
\end{align}

We investigate the bound of $G_3$ as follows,
\begin{equation}
    G_3 = \|\Theta_t-\Theta^*\|^2 \underbrace{- 2 \eta_t \langle \Theta_t-\Theta^*, \bar{f}_t \rangle}_{G_6}+ \eta_t^2 \|\bar{f}_t\|^2. \label{eq:lemma2-bound-B3}
\end{equation}

The term $G_6 / (2\eta_t)$ is bounded as,
\begin{eqnarray}
\frac{G_6}{2\eta_t} &\leqwhy{a}&
F(\Theta^*) - F(\Theta_t) - \frac{\mu}{2} \|\Theta_t-\Theta^*\|^2 \\
&\leqwhy{b}&  - \frac{1}{2\beta}\|\bar{f}_t\|^2- \frac{\mu}{2} \|\Theta_t-\Theta^*\|^2 \\ &\leqwhy{c}&  - \frac{\mu}{2} \|\Theta_t-\Theta^*\|^2. \label{eq:lemma2-B7}
\end{eqnarray}
The steps (a), (b) and (c) are due to $\mu$-strong convexity, $L$-smoothness, and $\|\bar{f}_t\|^2 \geq 0$, respectively.
Since $\mathbb{E}[f_t] =\bar{f}_t$, $\mathbb{E}[G_5] = 0$. 
Combining Lemma~\ref{lemma:2}, Lemma~\ref{lemma:3}, and these results, we have the bound of LHS of \eqref{eq:lem3-prob}. Summarizing \eqref{eq:lem2-bound} with taking expectation, 
and under Assumption~\ref{assum:1} with a learning rate $\eta_t \leq \frac{1}{\beta}$, the error between the updated global model and its optimum progress as,
\begin{align}
     \mathbb{E}&\|\Theta_{t+1}-\Theta^*\|^2  \leq (1-\eta_t\mu)\mathbb{E}\|\Theta_t-\Theta^*\|^2\nonumber\\ 
     & + \eta_t^2 \underbrace{\Big(EL^2(2-\lambda)^2 +  L\delta\Big) \sum^L_{l=1} \frac{1}{p_{l}^2} }_{:=B}. \label{eq:lemma3}
\end{align}

Since $\eta_t = \frac{2}{\mu{t}+2L-\mu} \leq \frac{1}{L}$, applying \eqref{eq:lemma3}, we have %$\Delta_{t+1} \leq (1-\eta_t\mu)\Delta_{t} + \eta_t^2 B$.
\begin{equation}
    \Delta_{t+1} \leq (1-\eta_t\mu)\Delta_{t} + \eta_t^2 B.
\end{equation}

For diminishing the step-size, we focus on showing that $\Delta_t \leq \frac{v}{t + 2\kappa-1}$, where $\kappa = \frac{\beta}{\mu}$ and $v = \max\{2\kappa\Delta_{1},{4B}/{\mu^2}\}$ as elaborated next.
It is trivial that $\Delta_1\leq\frac{v}{2\kappa}$ due to the definition of $v$. Assuming $\Delta_{t} \leq \frac{v}{t + 2\kappa-1}$, we have
    % $\Delta_{t+1} \leq (1-\mu\eta_{t})\Delta_{t} + \eta^{2}_{t}B
    % \leq \left(1- \frac{2}{t+2\kappa -1}\right) \frac{v}{t+2\kappa-1}+\frac{{4B}/{\mu^2}}{(t+2\kappa-1)^2}
    % = \frac{(t+2\kappa-2)v-(v-{4B}/{\mu^{2}})}{(t+2\kappa-1)^{2}} \leq \frac{t+2\kappa-2}{(t+2\kappa-1)^{2}}v 
    % \leq  \frac{v}{t+2\kappa}$.
\begin{align}
   &\Delta_{t+1} \leq (1-\mu\eta_{t})\Delta_{t} + \eta^{2}_{t}B \\
   &\leq \left(1- \frac{2}{t+2\kappa -1}\right) \frac{v}{t+2\kappa-1}+\frac{{4B}/{\mu^2}}{(t+2\kappa-1)^2} \\
   &= \frac{(t+2\kappa-2)v-(v-{4B}/{\mu^{2}})}{(t+2\kappa-1)^{2}} \leq \frac{t+2\kappa-2}{(t+2\kappa-1)^{2}}v \\
   &\leq  \frac{v}{t+2\kappa}.
\end{align}

For $t=1$, we obtain 
%$v = \max\{2\kappa\Delta_{1},\frac{4B}{\mu^2}\} \leq 2\kappa\Delta_{1} + \frac{4B}{\mu^2}$.
\begin{equation}
    v = \max\{2\kappa\Delta_{1},\frac{4B}{\mu^2}\} \leq 2\kappa\Delta_{1} + \frac{4B}{\mu^2}.
\end{equation}

Finally, using Assumption~\ref{assum:1}, \eqref{eq:lemma3}, the result above, we complete the proof of the theorem.
\bibliographystyle{IEEEtran}
% \balance 
\bibliography{refs}

\newpage

\begin{IEEEbiography}[{\includegraphics[width=1in,height=1.25in,clip]{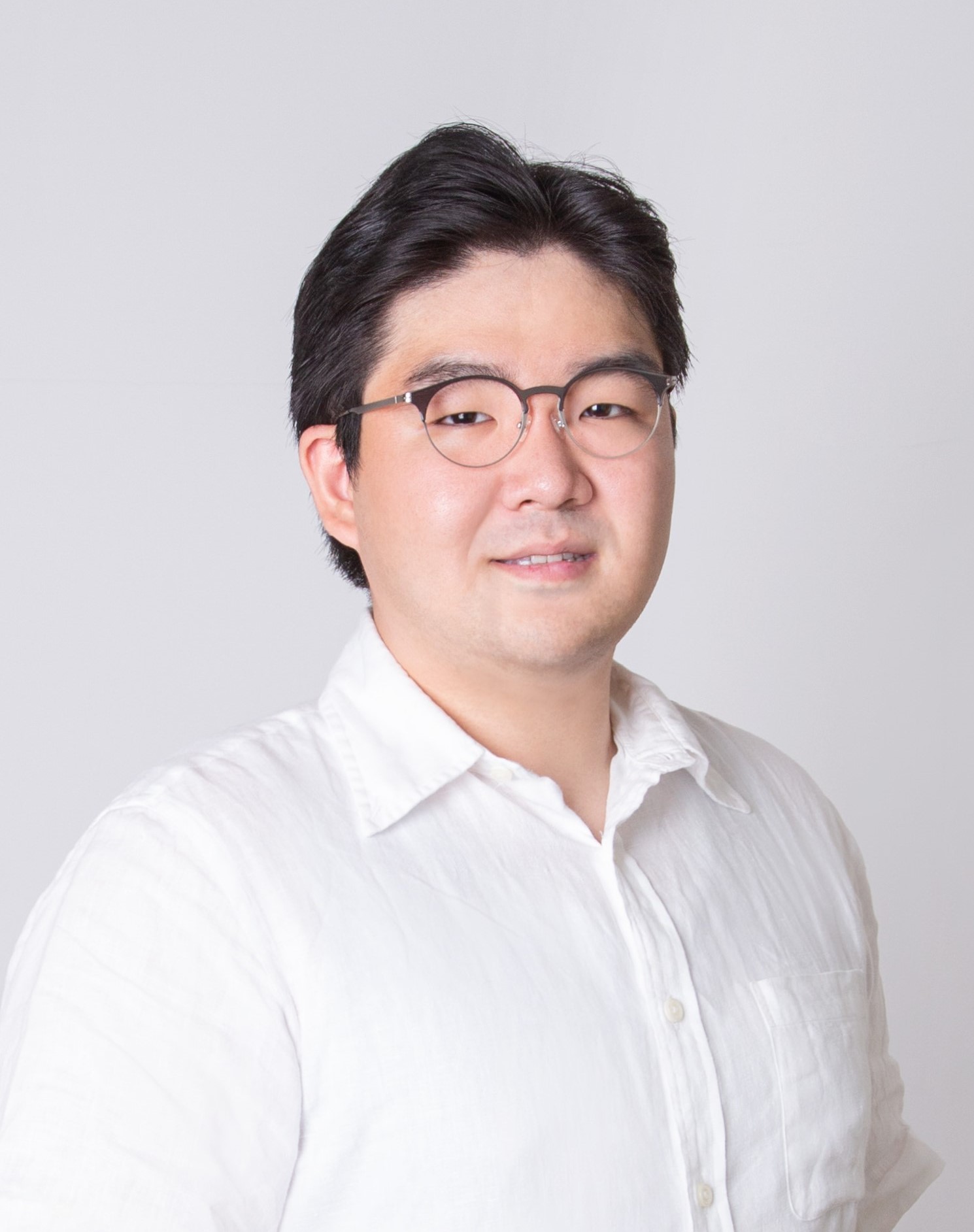}}]{Won Joon Yun} 
 is currently a Ph.D. student in electrical and computer engineering at Korea University, Seoul, Republic of Korea, since March 2021, where he received his B.S. in electrical engineering. He was a visiting researcher at Cipherome Inc., San Jose, CA, USA, during summer 2022; and also a visiting researcher at the University of Southern California, Los Angeles, CA, USA during winter 2022 for a joint project with Prof. Andreas F. Molisch at the Ming Hsieh Department of Electrical and Computer Engineering, USC Viterbi School of Engineering. 
\end{IEEEbiography}

\begin{IEEEbiography}[{\includegraphics[width=1in,height=1.25in,clip]{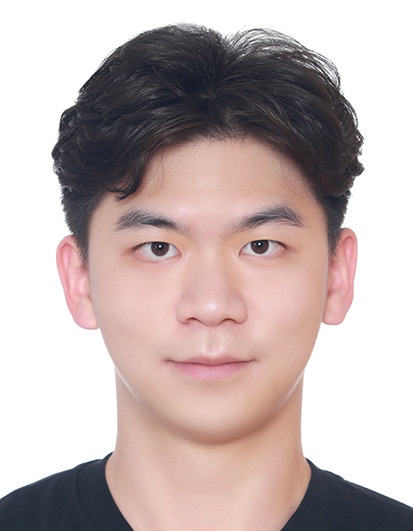}}]{Jae Pyoung Kim} has been with the School of Electrical Engineering, Korea University, Seoul, Republic of Korea, since March 2017, where he is currently a B.S. student in electrical and computer engineering. He is now a research engineer at Artificial Intelligence and Mobility (AIM) Laboratory at Korea University, Seoul, Republic of Korea, since 2021. 

His current research interests include quantum machine learning. 
\end{IEEEbiography}

\begin{IEEEbiography}[{\includegraphics[width=1in,height=1.25in,clip]{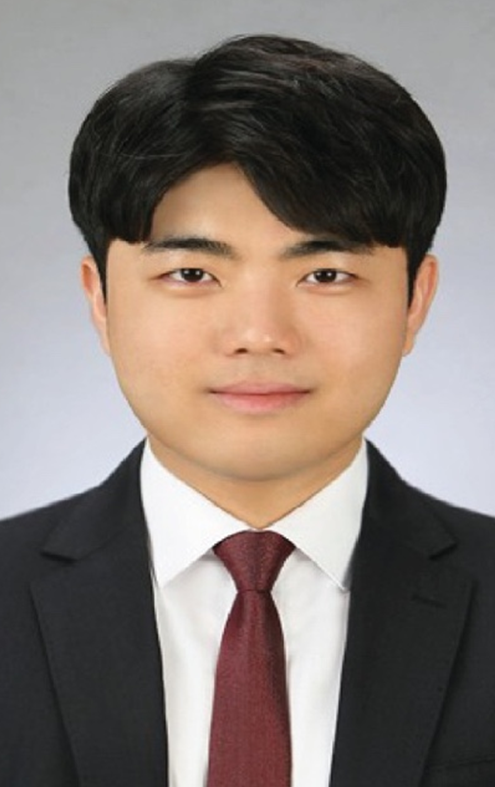}}]{Hankyul Baek} 
is currently a Ph.D. student in electrical and computer engineering at Korea University, Seoul, Republic of Korea, since March 2021. He received his B.S. in electrical engineering from Korea University, Seoul, Republic of Korea, in 2020. He was with LG Electronics, Seoul, Republic of Korea, from 2020 to 2021. 

His current research interests include quantum machine learning and its applications. 
\end{IEEEbiography}

\begin{IEEEbiography}[{\includegraphics[width=1in,height=1.25in,clip]{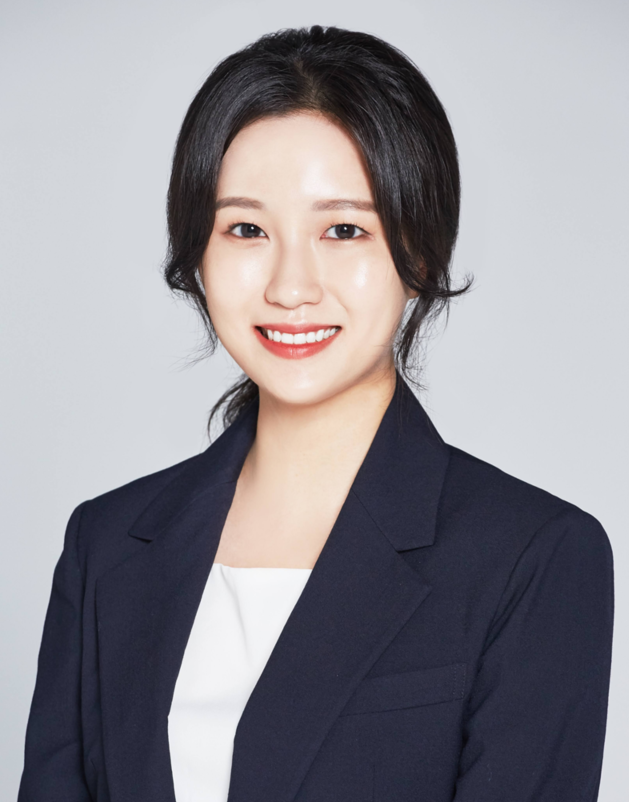}}]{Soyi Jung} has been an assistant professor at the department of electrical and computer engineering, Ajou University, Suwon, Republic of Korea, since September 2022. She also holds a visiting scholar position at Donald Bren School of Information and Computer Sciences, University of California, Irvine, CA, USA, from 2021 to 2022. She was a research professor at Korea University, Seoul, Republic of Korea, during 2021. She was also a researcher at Korea Testing and Research (KTR) Institute, Gwacheon, Republic of Korea, from 2015 to 2016. 
She received her B.S., M.S., and Ph.D. degrees in electrical and computer engineering from Ajou University, Suwon, Republic of Korea, in 2013, 2015, and 2021, respectively. 

Her current research interests include network optimization for autonomous vehicles communications, distributed system analysis, big-data processing platforms, and probabilistic access analysis. She was a recipient of Best Paper Award by KICS (2015), Young Women Researcher Award by WISET and KICS (2015), Bronze Paper Award from IEEE Seoul Section Student Paper Contest (2018), ICT Paper Contest Award by Electronic Times (2019), and IEEE ICOIN Best Paper Award (2021).
\end{IEEEbiography}

\begin{IEEEbiography}[{\includegraphics[width=1in,height=1.25in,clip]{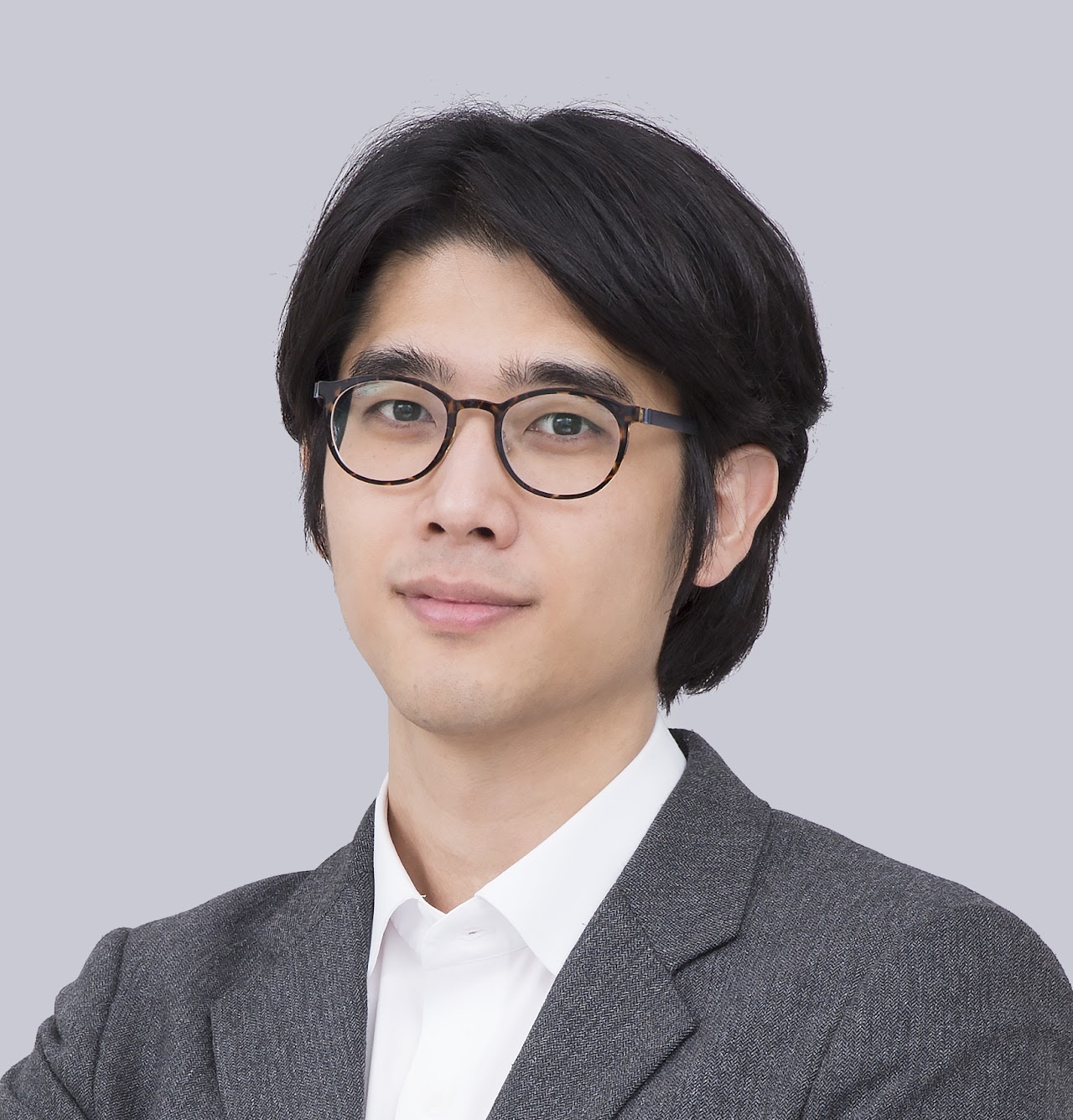}}]{Jihong Park} 
 (Senior Member, IEEE) received the B.S. and Ph.D. degrees from Yonsei University, South Korea. He is currently a Lecturer (Assistant Professor) with the School of Information Theory, Deakin University, Australia. 
 His research interests include ultra-dense/ultra-reliable/mmWave system designs, and distributed learning/control/ledger technologies and their applications for beyond-5G/6G communication systems. He served as a Conference/Workshop Program Committee Member for IEEE GLOBECOM, ICC, and WCNC, and for NeurIPS, ICML, and IJCAI. He is an Associate Editor of Frontiers in Data Science for Communications, and a Review Editor of Frontiers in Aerial and Space Networks.
\end{IEEEbiography}

\begin{IEEEbiography}[{\includegraphics[width=1in,height=1.25in,clip]{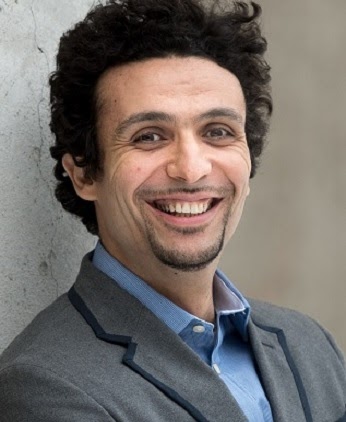}}]{Mehdi Bennis} 
 (Fellow, IEEE) is a tenured Full Professor with the Centre for Wireless Communications, University of Oulu, Finland, an Academy of Finland Research Fellow, and the Head of the Intelligent Connectivity and Networks/Systems Group (ICON). He has published more than 200 research papers in international conferences, journals, and book chapters. His main research interests are in radio resource management, heterogeneous networks, game theory, and distributed machine learning in 5G networks and beyond. 

 He has been the recipient of several prestigious awards, including the 2015 Fred W. Ellersick Prize from the IEEE Communications Society, the 2016 Best Tutorial Prize from the IEEE Communications Society, the 2017 EURASIP Best Paper Award for the Journal of Wireless Communications and Networks, the All-University of Oulu Award for research, the 2019 IEEE ComSoc Radio Communications Committee Early Achievement Award, and the 2020 Clarivate Highly Cited Researcher by the Web of Science. He is an Editor of IEEE Transactions on Communications and the Specialty Chief Editor of Data Science for Communications in the Frontiers in Communications and Networks.
\end{IEEEbiography}

\begin{IEEEbiography}[{\includegraphics[width=1in,height=1.25in,clip]{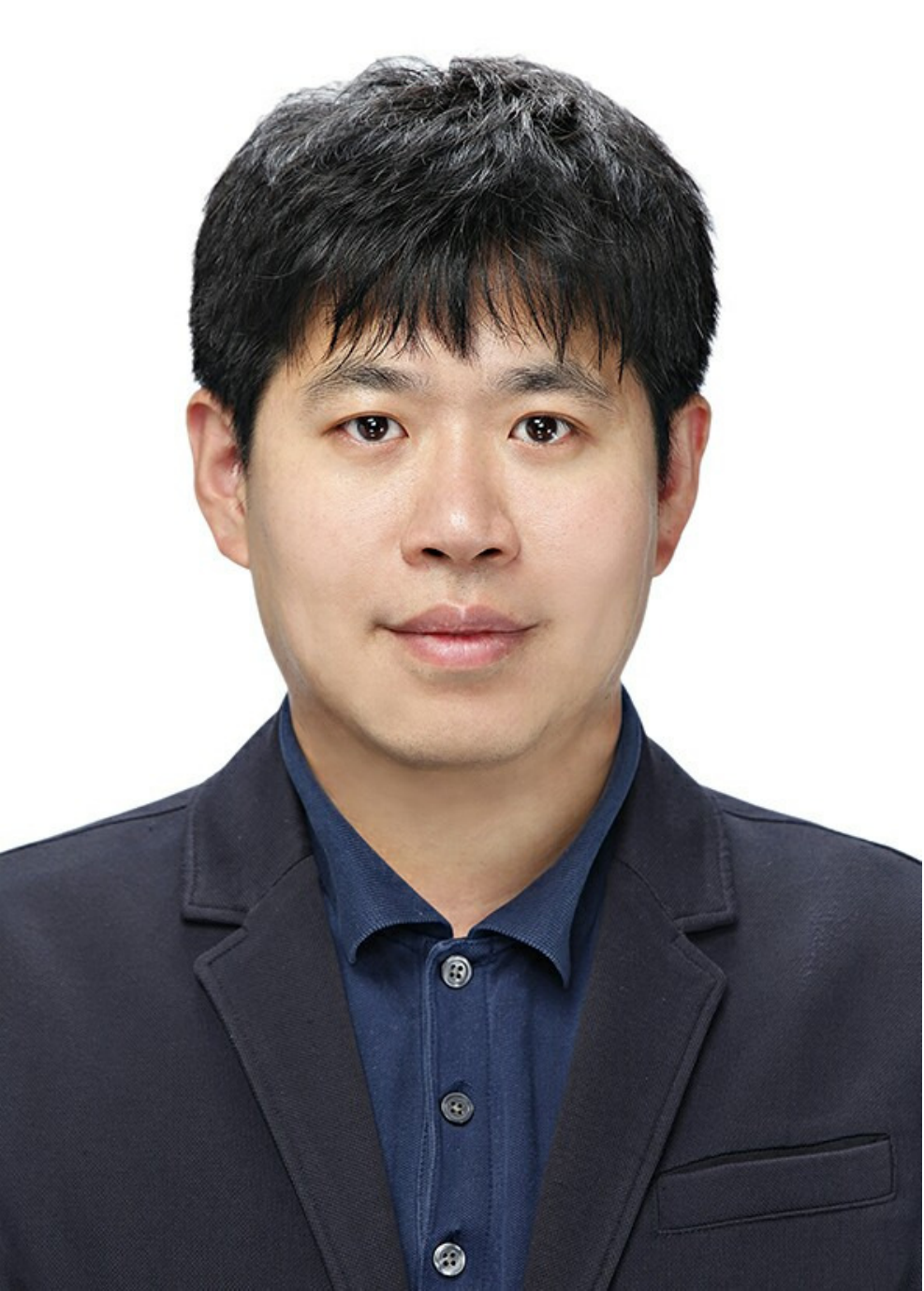}}]{Joongheon Kim}
(Senior Member, IEEE) has been with Korea University, Seoul, Korea, since 2019, where he is currently an associate professor. He received the B.S. and M.S. degrees in computer science and engineering from Korea University, Seoul, Korea, in 2004 and 2006, respectively; and the Ph.D. degree in computer science from the University of Southern California (USC), Los Angeles, CA, USA, in 2014. Before joining Korea University, he was with LG Electronics (Seoul, Korea, 2006--2009), Intel Corporation (Santa Clara in Silicon Valley, CA, USA, 2013--2016), and Chung-Ang University (Seoul, Korea, 2016--2019). 

 He serves as an editor for \textsc{IEEE Transactions on Vehicular Technology}, \textsc{IEEE Transactions on Machine Learning in Communications and Networking}, \textsc{IEEE Communications Standards Magazine}, \textit{Computer Networks (Elsevier)}, and \textit{ICT Express (Elsevier)}. He is also a distinguished lecturer for \textit{IEEE Communications Society (ComSoc)} (2022-2023) and \textit{IEEE Systems Council} (2022-2024).

 He was a recipient of Annenberg Graduate Fellowship with his Ph.D. admission from USC (2009), Intel Corporation Next Generation and Standards (NGS) Division Recognition Award (2015), \textsc{IEEE Systems Journal} Best Paper Award (2020), IEEE ComSoc Multimedia Communications Technical Committee (MMTC) Outstanding Young Researcher Award (2020), IEEE ComSoc MMTC Best Journal Paper Award (2021), and Best Special Issue Guest Editor Award by \textit{ICT Express (Elsevier)} (2022). He also received several awards from IEEE conferences including IEEE ICOIN Best Paper Award (2021), IEEE Vehicular Technology Society (VTS) Seoul Chapter Awards (2019, 2021), and IEEE ICTC Best Paper Award (2022). 
\end{IEEEbiography}
\end{document}